\documentclass[12pt]{article}

\input{epsf}
\usepackage[nosort]{cite}
\usepackage{cite}
\usepackage{amsmath,pifont,bbding}
\usepackage{epsfig}
\usepackage{amsfonts}
\usepackage{amssymb}
\usepackage{multirow}
\usepackage{enumerate}
\usepackage{subfigure}
\usepackage{slashed}
\usepackage{color}

\usepackage{subfigure}

\usepackage{caption}

\newcommand{\be}{\begin{equation}}
\newcommand{\ee}{\end{equation}}
\newcommand{\bee}{\begin{equation*}}
\newcommand{\eee}{\end{equation*}}
\newcommand{\bea}{\begin{eqnarray}}
\newcommand{\eea}{\end{eqnarray}}
\newcommand{\bean}{\begin{eqnarray*}}
\newcommand{\eean}{\end{eqnarray*}}

\begin{document}

\setcounter{page}{0}
\thispagestyle{empty}

\begin{flushright}
DESY 17-021
\end{flushright}

\vskip 8pt

\begin{center}
{\bf \Large {
Gravitational waves from the asymmetric-dark-matter generating phase transition
}}
\end{center}

\vskip 12pt

\begin{center}
 {\bf  Iason Baldes}
 \end{center}

\vskip 14pt

\begin{center}
\centerline{{\it DESY, Notkestra{\ss}e 85, D-22607 Hamburg, Germany}}

\vskip .3cm
\centerline{\tt iason.baldes@desy.de}
\end{center}

\vskip 10pt

\begin{abstract}
\vskip 3pt
\noindent
The baryon asymmetry, together with a dark matter asymmetry, may be produced during a first order phase transition in a generative sector. We study the possibility of a gravitational wave signal in a model realising such a scenario. We identify areas of parameter space with strong phase transitions which can be probed by future, space based, gravitational wave detectors. Other signals of this scenario include collider signatures of a $Z'$, DM self interactions, a contribution to $\Delta N_{\rm eff}$ and nuclear recoils at direct detection experiments. 

\end{abstract}

\newpage

\tableofcontents

\vskip 13pt


\section{Introduction}
There is overwhelming evidence for an exotic, non-baryonic, component to the cosmic matter density. However, the detailed nature of this dark matter is so far unknown. A possibility is for the DM density to be set in a similar way as the baryonic density~\cite{Nussinov:1985xr,Davoudiasl:2012uw,Petraki:2013wwa,Zurek:2013wia}. An asymmetry between particle and antiparticle number density is first  created in a dynamical process of baryogenesis~\cite{Sakharov:1967dj}. The symmetric component of the particle and antiparticle population is then efficiently annihilated into radiation degrees-of-freedom whose energy is redshifted away by the Hubble expansion~\cite{Chiu:1966kg}. The asymmetric component is left over, which forms the matter density observed today.

Various methods to produce an asymmetry in both the dark and visible sectors concurrently have been proposed. The focus of this paper will be a model proposed by Petraki, Trodden and Volkas~\cite{Petraki:2011mv}, in which the asymmetries in the dark and visible sectors are produced during a strong first order phase transition, in analogy with the electroweak baryogenesis mechanism~\cite{Kuzmin:1985mm,Shaposhnikov:1986jp,Shaposhnikov:1987tw,Cohen:1993nk,Trodden:1998ym,Morrissey:2012db}. While we focus on a particular realisation, the possibility of observational gravitational waves in such models is generic. This is because they all rely on beyond-the-standard-model (BSM) chiral fermions undergoing CP violating interactions with the broken phase bubbles of an exotic non-abelian gauge group in order to generate the asymmetry~\cite{Dutta:2006pt,Shelton:2010ta,Dutta:2010va,Walker:2012ka,Davoudiasl:2016ijt}. The presence of a non-minimal dark sector is also generic in such models. Since~\cite{Petraki:2011mv} appeared, much progress has been made on both the experimental and theoretical fronts, in physics relevant to the phenomenology of such models. 

On the experimental side, a SM-like Higgs boson has been discovered and studied at the LHC~\cite{Chatrchyan:2012xdj,Aad:2012tfa}. The same machine has led to improved limits being placed on $Z'$ bosons~\cite{ATLAS-CONF-2016-045,Khachatryan:2016zqb}. Planck has set improved limits on exotic radiation present during the CMB ($\Delta N_{\rm eff}$)~\cite{Ade:2015xua}. LIGO has detected gravitational waves~\cite{Abbott:2016blz,Abbott:2016nmj} and the LISA pathfinder has flown which returned promising results on the prospects for space-based gravitational wave interferometers~\cite{Armano:2016bkm}. On the theoretical side, simulations of the sound waves produced in the thermal plasma during a first order phase transition have found these can be a strong source of gravitational waves, which improves the possibility of such a  detection~\cite{Hindmarsh:2013xza,Hindmarsh:2015qta,Hindmarsh:2016lnk}.

Production of gravitational waves by exotic phase transitions has been discussed in~\cite{Grojean:2006bp,Buchmuller:2013lra,Schwaller:2015tja,Jaeckel:2016jlh,Dev:2016feu,Davoudiasl:2016yfa,Kobakhidze:2016mch,GarciaGarcia:2016xgv,Addazi:2016fbj,Katz:2016adq,Balazs:2016tbi,Huang:2017laj}. A phase transition in a \emph{generative} sector, producing an asymmetry which is then communicated to the dark and visible matter sectors, is one such possibility. Unlike the electroweak sector, the exotic generative sector Higgs mass, VEV, gauge and fermionic couplings have not been observed. Hence there is a greater range of possibilities as to the mass scale of such a sector and therefore also more possibilities for the phase transition temperature and peak gravitational wave frequency. The CP violation required in these scenarios occurs in the generative sector Yukawa interactions and so EDM constraints are easily satisfied~\cite{Baker:2006ts,Hudson:2011zz,Baron:2013eja,Chao:2014dpa,Dorsch:2016nrg,Balazs:2016yvi}.

It is therefore timely to revisit the model proposed in~\cite{Petraki:2011mv}. In section~2 we briefly review the model. In section~3 we present the main investigation of this paper, we calculate the strength of the phase transition, which allows us to find the stochastic gravitational wave background in terms of the underlying parameters of the model.\footnote{The goal of \cite{Petraki:2011mv} was to construct a concrete model realising asymmetric DM via a first order phase transition. Once the field content was specified, it was clear, in analogy with the studies of EW baryogenesis, that a strong first order phase transition was possible. Hence the details of the phase transition were irrelevant to the aims of \cite{Petraki:2011mv}. However, in the present work, in which the gravitational wave signal is a focus, the details of the phase transition are of central importance.} In section~4 we investigate other relevant signals, i.e. $Z_{\rm B-L}'$ at the LHC, constraints from halo ellipticity due to DM self interactions, $\Delta N_{\rm eff}$ and direct detection prospects. We then briefly discuss and conclude.

\section{Review of the model}
We now give an overview of the model presented in~\cite{Petraki:2011mv}. The symmetry structure of the model includes three new gauge groups and an anomalous global symmetry $U(1)_{X}$. The field content is summarised in table~\ref{tab:fieldcontent}. The model may be organised into three sectors.

\begin{table}[t]
\begin{center}
\begin{tabular}{|c||l|c|c|c|c|}
\hline 
\multirow{2}{*}{Sector} & \multirow{2}{*}{Particles}  & SU(2)$_G$ & U(1)$_{B-L}$ & U(1)$_D$	& U(1)$_X$       \\
                         &                            & (gauged)  & (gauged)     & (gauged)     & (anomalous)    \\  
\hline \hline
                         & $\Psi_L$                   &  2        &  0        	 & 0  	        & -2             \\
Generative               & $\Psi_{1R}, \Psi_{2R}$     &  1        &  0           & 0	        & -2             \\ 
                         & $\phi$                     &  2        &  0           & 0	        &  0             \\ \hline
\multirow{2}{*}{Visible} & $ f_{L,R}$                 &  1        & -1           & 0	        & -1             \\ 
                         & $ \nu_R$                   &  1        & -1           & 0            & -1             \\ \hline
                         & $\chi$                     &  2        &  1       	 & 0 	        & -1             \\ 
Dark                     & $\xi_{L,R}$                &  2        &  0           & 1            &  0             \\ 
                         & $\zeta_{L,R}$              &  1        & -1         	 & 1 	        &  1             \\ \hline
$B-L$  Higgs	         &  $\sigma$                  &  0        &  $q^{\sigma}_{B-L}$  & {0}          &  0	   \\ 
\hline
\end{tabular}
\end{center}
\caption{The field content and charges of the model. The three right-handed neutrinos are introduced to cancel the cubic $B-L$ anomaly. The $SU(2)_G$ and $U(1)_{B-L}$ symmetries are broken spontaneously at a high, $\mathcal{O}($TeV), scale. The $U(1)_D$ symmetry can either remain exact or be broken spontaneously at a suitably low scale, to allow the $\mathcal{O}$(GeV) scale DM to annihilate into dark photons.
}
\label{tab:fieldcontent}
\end{table}

\begin{enumerate}[(i)]
\item The \emph{generative} sector, which consists of chiral fermions, $\Psi_{L}$, $\Psi_{1R}$, $\Psi_{2R}$, and a generative sector Higgs, $\phi$. This Higgs breaks the generative sector $SU(2)_{G}$ gauge group and generates mass for the sector's chiral fermions through Yukawa terms of the form (for one generation)
	\begin{equation}
	\mathcal{L} \supset -\frac{1}{\sqrt{2}}\left(\sum_{j=1}^{2}h_{j}\overline{\Psi_{L}}\phi\Psi_{jR}+\tilde{h}_{j}\overline{\Psi_{L}}\tilde{\phi}\Psi_{jR} \right)+H.c.
	\label{eq:genyuk}
	\end{equation}
We require an even number of generations of $\Psi$ fermions to be free of the Witten anomaly~\cite{Witten:1982fp}. In our calculations below we always take the minimal choice of two generations of $\Psi$ fermions corresponding to 16 degrees of freedom. The Yukawa couplings of eq.~(\ref{eq:genyuk}) can be complex and after rotating to the mass basis, we find the choice of two generations gives us seven physical CP violating phases, so that the second Sakharov condition is fulfilled.\footnote{There are more phases present than in the SM case, which requires three generations of quarks for one CP violating phase, because the absence of hypercharge allows both upper and lower $\Psi_{L}$ components a separate mass term with $\Psi_{R1}$ and $\Psi_{R2}$ in eq.~(\ref{eq:genyuk}). Majorana mass terms are prohibited by imposing the global $U(1)_{X}$ symmetry.} The fermions carry charge under an anomalous $U(1)_X$, which is violated by rapid $SU(2)_{G}$ sphalerons in the symmetric phase. CP violating interactions of the fermions with the bubble walls of broken $SU(2)_{G}$ phase during a first order phase transition can lead to an $U(1)_{X}$ asymmetry forming outside the bubble walls. This asymmetry is swept into the expanding bubble in which, if the phase transition is strong enough, the $SU(2)_{G}$ sphalerons are suppressed and the asymmetry is frozen in. The generated asymmetry is transfered to the visible and dark sectors through a chain of Yukawa interactions beginning with
	\begin{equation}
	\mathcal{L} \supset -\frac{\kappa}{\sqrt{2}}\overline{\Psi_{L}}\chi f_{R} +H.c.
	\end{equation}
where the fermion $f_{R}$ (scalar $\chi$) will communicate with the visible (dark) sector, as explained below.
\item The \emph{visible} sector consists of the SM together with vector-like fermions, $f_{L}$ and $f_{R}$, neutral under the SM gauge group. These transfer the asymmetry, created in the $\Psi$'s during the generative phase transition, into the SM via the interactions
	\begin{equation}
	\mathcal{L} \supset -\frac{y_{1}}{\sqrt{2}}\overline{l_{L}}Hf_{R} + H.c.
	\end{equation}
where $l_{L}$ ($H$) is the SM lepton (Higgs) doublet.
\item The \emph{dark} sector consists of the scalar $\chi$ together with vector-like fermions $\xi_{L,R}$ and $\zeta_{L,R}$ which make up the DM. The asymmetry is transfered via the interaction
	\begin{equation}
	\mathcal{L} \supset -\frac{y_{2}}{\sqrt{2}}\overline{\xi}\chi\zeta + H.c.
	\end{equation}
The scalar $\chi$ retains a zero VEV. The fermions $\xi_{L,R}$ and $\zeta_{L,R}$ are charged under a local $U(1)_{D}$ and may form atomic bound states if the symmetry remains unbroken or is mildly broken. The symmetric components of the $\xi_{L,R}$ and $\zeta_{L,R}$ number densities are efficiently annihilated away into dark $U(1)_{D}$ photons. The $U(1)_{D}$ gauge symmetry may either remain unbroken, or be broken spontaneously at a suitably low scale, in order to still allow for efficient annihilation of the dark fermions which have masses $m_{\zeta}+m_{\chi}\approx 1.5$ GeV.  
\end{enumerate}

The above review of the model suffices for the casual reader interested in the discussion below. More discussion is provided in~\cite{Petraki:2011mv} for those interested in understanding the in-depth details of the model.

\section{Phase transition and gravitational wave signal}
\subsection{The potential}
In order to study the strength of the phase transition we must make some assumptions about the tree level potential. Here we assume terms up to dimension six term in the tree level potential~\cite{Grojean:2004xa,Bodeker:2004ws,Delaunay:2007wb}
	\begin{equation}
	V_{G} = \frac{ \mu_{\phi}^{2} }{ 2 }\phi^{2} + \frac{ \lambda_{ \phi }}{4} \phi^{4} + \frac{1}{8\Lambda_{\phi}^{2}}\phi^{6}.
	\label{eq:v6pot}
	\end{equation}
Such a potential can correspond to the low energy effective theory if there is a heavier scalar to which $\phi$ couples sufficiently strongly~\cite{Damgaard:2015con}.\footnote{The simpler possibility of including only terms up to $\phi^{4}$ is relegated to appendix A, as its finite temperature perturbative expansion suffers pathological issues due to the large gauge couplings required in order to achieve a sufficiently strong first order phase transition for an appreciable gravitational wave signal.} Taking $\mu_{\phi}, \; \Lambda_{\phi} > 0$ and $\lambda_{\phi} < 0$, the VEV is given by
	\begin{equation}
	v_{\phi}^{2}=\frac{2\Lambda_{\phi}^{2}}{3}\left(-\lambda_{ \phi } + \sqrt{\lambda_{ \phi }^{2} - 3\frac{\mu_{\phi}^{2}}{\Lambda_{\phi}^{2}}}\right),
	\end{equation}
and the field dependent Higgs mass by
	\begin{equation}
	m_{\phi}^{2}=\mu_{\phi}^{2} + 3\lambda_{ \phi }\phi^2 + \frac{15}{4\Lambda_{\phi}^{2}}\phi^4.
	\end{equation}
The field dependent masses of the $\Psi$ fermions due to the Yukawa interactions of eq.~(\ref{eq:genyuk}) and the $SU(2)_{G}$ gauge bosons are given by
	\begin{equation}
	\begin{gathered}
	m_{\Psi}  =\frac{h}{\sqrt{2}}\phi, \qquad m_{G}  = \frac{g_{G}}{2}\phi,
	\end{gathered}
	\end{equation}
where $g_{G}$ is the $SU(2)_{G}$ gauge coupling and we assume a universal, representative, Yukawa coupling in our calculations for simplicity. In order to analyse the strength of the phase transition we use the effective potential, $V_{\rm eff}$, at one loop order including $T=0$ and finite temperature loop contributions of $\phi$, the $SU(2)_{G}$ gauge bosons and two generations of $\Psi$ fermions~\cite{Quiros:1999jp}. The daisy correction is also included. The details can be found in appendix~B. The integrals which enter the effective potential are evaluated numerically.

\begin{figure}[t]
\begin{center}
\includegraphics[width=230pt]{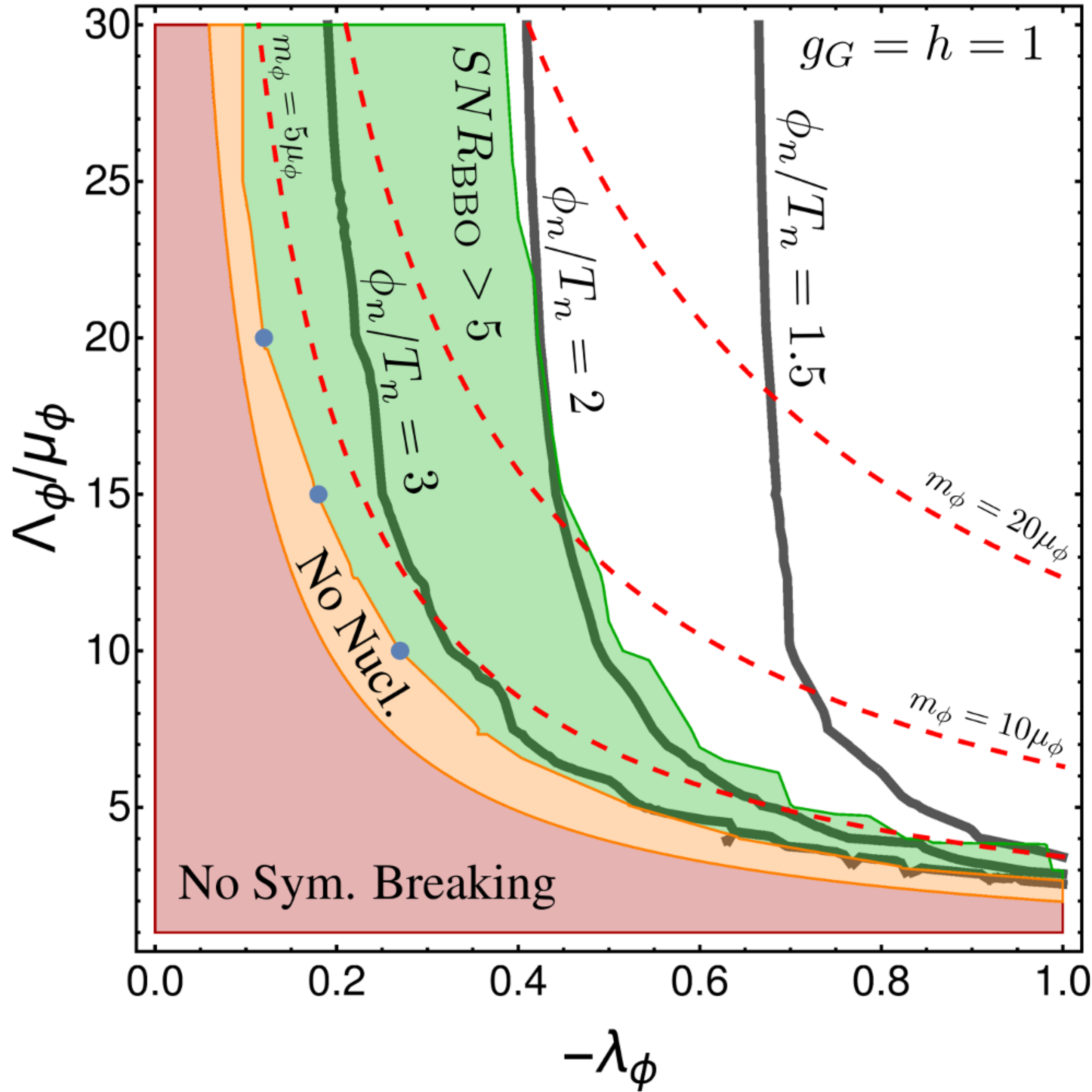}
\end{center}
\caption{Contours of $\phi_{n}/T_{n}$ (thick grey lines) for the $\phi^{6}$ tree level potential of eq.~(\ref{eq:v6pot}). The washout condition of eq.~(\ref{eq:washoutcondition}) is satisfied for $\phi_{n}/T_{n}\gtrsim 1.5 - 1.8$. The red area does not lead to symmetry breaking as $V_{\rm eff}^{T=0}(v_{\phi})>V_{\rm eff}^{T=0}(0)$. The orange region is either meta-stable and does not lead to a thermal transition or it leads to a runaway transition. The green area returns a gravitational wave spectrum with a BBO signal to noise ratio (SNR) $>5$. The SNR for gravitational waves has been calculated with $\mu_{\phi}=100$ GeV, while all other contours are independent of $\mu_{\phi}$, see sections~\ref{sec:gw} and \ref{sec:scaling} for details. Note we have used an optimistic value for the wall velocity, $v_{w}=1/\sqrt{3}$, in calculating the SNR. The blue points are where we also find a LISA SNR  $>5$. These always occur close to runaway transitions in this model. Hence LISA is only sensitive to a very small portion of the parameter space of this model.}
\label{fig:phi6_scan}
\end{figure}

\subsection{Washout condition}
A first order phase transition is characterised by a barrier in the effective potential, between the symmetric and broken phase minima, when the two have the same energy, i.e. at the critical temperature $T_{c}$. We denote the broken phase VEV at the critical temperature as $\phi_{c}$. Sometime after $T_{c}$ is reached, bubbles of broken phase can begin to nucleate. In order to avoid washout of the asymmetry, the $SU(2)_{G}$ sphalerons have to be sufficiently suppressed in the broken phase. This requires a sphaleron energy
	\begin{equation}
 	\frac{E_{\rm sph}(\phi_{n})}{T_{n}} = \frac{4\pi B}{g_{G}}\frac{\phi_{n}}{T_{n}} \gtrsim 37-45,
	\end{equation}
where $T_{n}$ is the bubble nucleation temperature, $\phi_{n}$ is the minimum of the potential at $T_{n}$, $B$ is an $\mathcal{O}(1)$ number which depends on the couplings of the theory and the range given on the right hand side of the inequality is due to the uncertainty in the sphaleron transition rate prefactor~\cite{Quiros:1999jp,Carson:1990jm}. Determining the $SU(2)_{G}$ sphaleron energy, i.e. calculating $B$, is beyond the scope of this work. In the SM, $B(\sqrt{\lambda}/g) \approx 1.58+0.91\sqrt{\lambda}/g-0.4\lambda/g$~\cite{Quiros:1999jp,Klinkhamer:1984di}. The addition to the SM of a higher dimensional Higgs term, $(H^{\dagger}H)^{3}/\Lambda^{2}$, decreases the sphaleron energy by only a few percent for $\Lambda\approx 500$ GeV~\cite{Grojean:2004xa}. Compared to the SM, the main difference to $E_{\rm sph}$ is therefore expected to come from $g_{G}$. The washout avoidance condition becomes
	\begin{equation}
	\label{eq:washoutcondition}
	\frac{\phi_{n}}{T_{n}} \gtrsim g_{G} \;  (1.5 - 1.8) \left( \frac{ 2.0 }{ B } \right).
	\end{equation}
As we increase the gauge coupling $g_{G}$ we therefore require a slightly larger $\phi_{n}/T_{n}$ ratio than is usually the case. 

As is well known, thermal one loop corrections from the gauge bosons and fermions lead to a $\phi^{2}T^{2}$ term in the effective potential, which makes the symmetric minimum the global one at high temperatures. As the temperature drops, with the potential of eq.~(\ref{eq:v6pot}), the phase transition can be strongly first order and satisfy the washout avoidance condition if the tree level barrier is large enough. An additional effect comes from the cancellation between the $\phi^{2}T^{2}$ term and the tree level $-\lambda_{\phi} \phi^{4}$ term, which can lead to a large barrier forming between the symmetric and broken phase minima. A scan over the parameters of the model showing contours of $\phi_{n}/T_{n}$ is shown in figure~\ref{fig:phi6_scan}.

As can be seen from the figure, for small $\Lambda_{\phi}/\mu_{\phi}$, the tree level barrier is important, while for large $\Lambda_{\phi}/\mu_{\phi}$, the cancellation between the $\phi^{2}T^{2}$ and $-\lambda_{\phi} \phi^{4}$ terms becomes the dominant effect. To better understand the somewhat counterintuitive behaviour for increasing $\Lambda_{\phi}$ at the top left of the figure, note that the VEV $v_{\phi}$ and mass $m_{\phi}$ have not been fixed. As $\Lambda_{\phi}$ increases so does $v_{\phi}$ --- together with $\phi_{n}$ and $T_{n}$ --- as the strength of the phase transition is then set largely by the cancellation between the $\phi^{2}T^{2}$ and $-\lambda_{\phi} \phi^{4}$ terms. We have checked that if $v_{\phi}$ and $m_{\phi}$ are instead held fixed, we recover the intuitive qualitative behaviour known from studies of the EW phase transition, in which increasing $\Lambda_{\phi}$ for fixed $v_{\phi}$ and $m_{\phi}$ results in a weaker phase transition~\cite{Grojean:2004xa,Delaunay:2007wb,Bodeker:2004ws}. Also shown in the figure are regions of possible observable gravitational waves. Details of the calculation are given in sections~\ref{sec:critbub}--\ref{sec:scaling}.

\subsection{Critical bubble}
\label{sec:critbub}
The first step in our calculation is to determine the nucleation temperature. The probability of creating a bubble within the Hubble volume, for $T \approx 0.1 -1$ TeV, reaches unity at
	\begin{equation}
	S_{3}/T \approx 140,
	\label{eq:nucleation}	
	\end{equation}
where $S_{3}$ is the three-dimensional Euclidean action for an O(3)-symmetric bubble~\cite{Quiros:1999jp,Linde:1980tt,Anderson:1991zb}. The temperature when this occurs is defined as the nucleation temperature, $T_{n}$, and $\phi_{n}$ is the minimum of the potential at $T_{n}$. The action is given by
	\begin{equation}
	S_{3}=4\pi\int r^{2} \left(\frac{1}{2}\left(\frac{d\phi}{dr}\right)^{2}+V_{\rm eff}\right)dr,
	\label{eq:S3}
	\end{equation}
evaluated for a $\phi(r)$ path with $\delta S_{3}=0$. The first term in the action corresponds to the cost of the surface tension, the second term to the favourability in terms of energy in residing in the deeper, broken phase, minimum. The resulting equation of motion is given by
	\begin{equation}
	\frac{d^{2}\phi}{dr^{2}}+\frac{2}{r}\frac{d\phi}{dr}=\frac{\partial V_{\rm eff}}{\partial \phi},
	\end{equation}
with the boundary conditions 
	\begin{align}
        \phi'(r=0)=0, \qquad \phi(r \to \infty)=0. \label{eq:rinfcond}
	\end{align}
We solve the equation of motion numerically. In practical terms, we start with some choice for the release point, $\phi(r=0)$, and then use an overshoot/undershoot method to determine the value $\phi(r=0)$ which returns $\phi(r \to \infty)=0$. One then substitutes the bubble profile, $\phi(r)$, one has found back into eq.~(\ref{eq:S3}), in order to determine $S_{3}/T$.  This allows us to calculate the contours of $\phi_{n}/T_{n}$ shown in figure~\ref{fig:phi6_scan}. Note that once the bubble begins to expand, the field value inside will approach the minimum of the potential at the relevant temperature, rather than $\phi(r=0)$, which motivates our use of the former as our definition of $\phi_{n}$. This choice makes negligible differences to figure~\ref{fig:phi6_scan}.

\subsection{Gravitational wave signal}
\label{sec:gw}

The dominant source of gravitational waves in this model comes from sound waves generated in the plasma by the bubbles. The frequency spectrum of gravitational waves,
	\begin{equation}
	h^{2}\Omega_{\rm GW}(f) \equiv h^{2}\frac{1}{\rho_{c}}\frac{d  \rho_{\rm GW}}{d \; \mathrm{Ln}f},
	\end{equation}
where $\rho_{\rm GW}$ is the energy density of gravitational waves and $\rho_{c}$ is the critical density, has been determined by fitting numerical simulations~\cite{Hindmarsh:2013xza,Hindmarsh:2015qta}. (An analytic approach has been pursued in~\cite{Hindmarsh:2016lnk}.) We give the key expressions below. An important quantity is related to how quickly the phase transition occurs
	\begin{equation}
	\label{eq:beta}
	\frac{\beta}{H} \equiv T_{n} \frac{d}{dT}\left(\frac{S_3}{T}\right)\bigg|_{T_n}.
	\end{equation}
Phase transitions which take longer result in larger, fewer, and more powerful bubbles, these produce sound waves at lower frequencies and with stronger amplitudes, than the smaller and more common bubbles found in faster transitions. The amplitude of gravitational waves also depends on the vacuum energy density released by the phase transition, compared with the radiation density of the thermal plasma,
	\begin{equation}
	\alpha \equiv \frac{\rho_{\rm vac}}{\rho_{\rm rad}} = \frac{\rho_{\rm vac}}{g_{\ast}\pi^{2}T_{n}^{4}/30},
	\label{eq:alpha}
	\end{equation}
where $g_{\ast}$ counts the effective number of relativistic degrees of freedom. Given the field content of this model, $g_{\ast}=165$ when all species are relativistic and we have used this value throughout our calculations. Using the method described to calculate the critical bubble, we not only find $T_{n}$, but also $\alpha$ and $\beta/H$. We also need the fraction of vacuum energy converted into bulk motion of the fluid, which for wall velocities $v_{w} \leq v_{\rm sound}$~\cite{Espinosa:2010hh}, is given by
	\begin{equation}
	\kappa_{v} = \frac{\rho_{v}}{\rho_{\rm vac}} \approx \frac{ v_{\rm sound}^{11/5}k_{A}k_{B} }{ (v_{\rm sound}^{11/5}-v_{w}^{11/5})k_{B} + v_{w}v_{\rm sound}^{6/5}k_{A} }
	\end{equation}
where
	\begin{align}
	k_{A} = \frac{6.9 \alpha v_{w}^{6/5}}{1.36-0.037\sqrt{\alpha}+\alpha}, \qquad
	k_{B} =  \frac{\alpha^{2/5}}{0.017+(0.997+\alpha)^{2/5}}.
	\end{align}
\begin{figure}[t]
\begin{center}
\includegraphics[width=300pt]{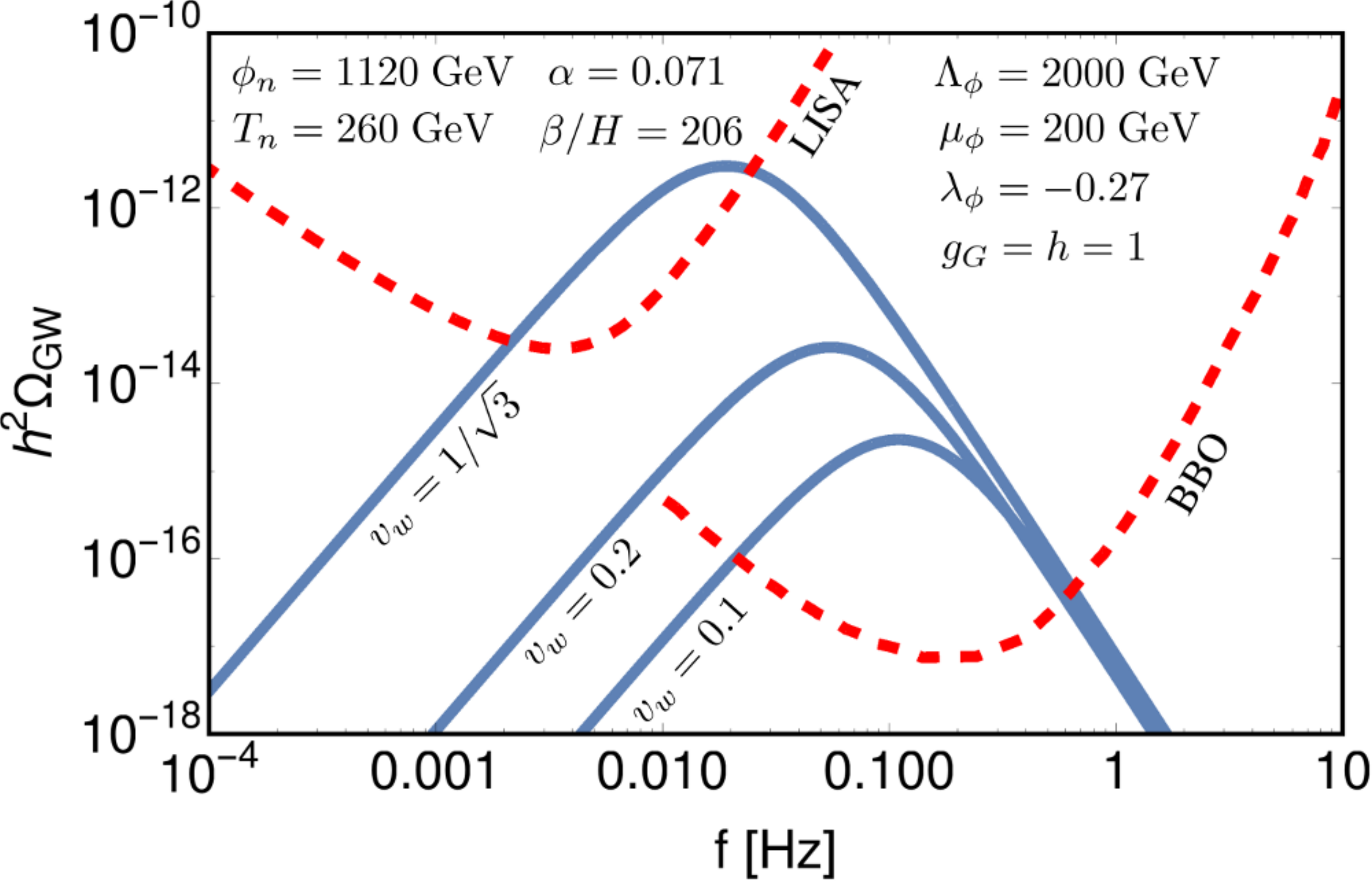}
\end{center}
\caption{Example of the gravitational wave spectrum for different wall velocities and estimated sensitivity of LISA (C1)~\cite{Caprini:2015zlo} and BBO~\cite{Thrane:2013oya} to stochastic power law backgrounds. The input parameters correspond to the lowest of the blue points in figure~\ref{fig:phi6_scan}. This point is close to the runaway boundary so a deflagration, with $v_{w} < 1/\sqrt{3}$ is actually unlikely, which illustrates the difficulty in obtaining a signal observable by LISA and consistent with baryogenesis in this model.}
\label{fig:gw}
\end{figure}
The numerical results are approximated by~\cite{Caprini:2015zlo}
	\begin{equation}
	h^{2}\Omega_{\rm GW}(f)=2.65\times 10^{-6} \left( \frac{H}{\beta} \right) \left( \frac{k_{v}\alpha}{1+\alpha} \right)^{2} \left( \frac{100}{g_{\ast}} \right)^{1/3}  S_{\rm sw}(f) \; v_{w},
	\end{equation}
where the spectral shape is
	\begin{equation}
	S_{\rm sw}(f) = \left(\frac{f}{f_{\rm sw}}\right)^{3} \left( \frac{7}{4+ 3(f/f_{\rm sw})^{2}} \right)^{7/2},
	\end{equation}
with peak frequency
	\begin{equation}
	\label{eq:peakfreq}
	f_{\rm sw} = 1.9 \times 10^{-5} \; \mathrm{Hz} \; \frac{1}{v_{w}} \left( \frac{\beta}{H} \right) \left( \frac{T_{n}}{100 \; \mathrm{GeV}} \right)  \left( \frac{g_{\ast}}{100} \right)^{1/6}.
	\end{equation}
The only missing ingredient required for the gravitational wave spectrum is the wall velocity, $v_{w}$. This depends on the strength of the phase transition and the friction of the plasma on the expanding bubble wall. This quantity is notoriously difficult to determine~\cite{Moore:1995si,Moore:2000wx,Megevand:2009gh,Konstandin:2010dm,Huber:2011aa,Huber:2013kj,Megevand:2013hwa,Kozaczuk:2014kva,Kozaczuk:2015owa}. We will use an optimistic value by setting $v_{w}=v_{\rm sound}=1/\sqrt{3}$ in our scan. This results in the largest gravitational wave amplitude possible while remaining consistent with electroweak-style baryogenesis, which requires subsonic wall velocities~\cite{No:2011fi}. The true $\Omega_{\rm GW}(f)$ will therefore be below our estimate (or the phase transition will be too strong for baryogenesis). Hence, our results should be seen as an optimistic estimate for the possibility of such a signal, until a detailed determination of $v_{w}$ can be performed for this scenario. In figure~\ref{fig:gw} we illustrate the sensitivity of the spectra to $v_{w}$.

Though we have not calculated the wall velocity, we can use a simple criterion to determine whether the phase transition is a runaway, in which $v_{w} \to 1$. The phase transition is necessarily a runaway if the mean field potential,
	\begin{equation}
	\overline{V} = V_{\rm tree}(\phi) + \frac{T^{2}}{24}\left(\sum_{\rm bosons}\left[ m_{b}^{2}(\phi)-m_{b}^{2}(0) \right]+\frac{1}{2}\sum_{\rm fermions}\left[m_{f}^{2}(\phi)-m_{f}^{2}(0) \right] \right),
	\end{equation}
drops below zero for $\phi \neq 0$ at $T_{n}$~\cite{Bodeker:2009qy}. If this is the case, the friction on the wall in the limit $v_{w} \to 1$, is insufficient to balance the pressure on the wall from the free energy difference between the symmetric and broken phases. Clearly, for consistency, we are interested in non-runaway phase transitions.

To produce gravitational waves observable at LISA and BBO requires relatively strong transitions. The detectability of the gravitational waves can be estimated by the signal to noise ratio~\cite{Caprini:2015zlo}
	\begin{equation}
	\mathrm{SNR} = \sqrt{t_{\rm obs} \int \left[\frac{h^{2}\Omega_{\rm GW}(f)}{h^{2}\Omega_{\rm sens}(f)}\right]^{2} df}, 
	\end{equation}
where $t_{\rm obs}$ is the time of observation in years and $h^{2}\Omega_{\rm sens}(f)$ is the sensitivity of the gravitational wave detector to stochastic backgrounds for one year of observation. Sensitivity curves for LISA and BBO for $t_{\rm obs}=1$ year are shown in figure~\ref{fig:gw}. Using $t_{\rm obs}=5$ years we show parameter space with SNR $>5$ for BBO in figure~\ref{fig:phi6_scan}. Note that due to the fermionic and gauge couplings even some phase transitions strong enough to enter the LISA sensitivity region can remain outside the runaway regime. However, only a small portion of the parameter space can be probed by LISA.\footnote{This conclusion does not hold if the phase transition is instead driven by nearly conformal dynamics. In such a case the phase transition can be generically strong enough for gravitational waves to be observable at LISA sensitivities~\cite{Creminelli:2001th,Nardini:2007me,Randall:2006py,Konstandin:2010cd,Konstandin:2011dr,Konstandin:2011ds}.} Another crucial issue for such searches will be the foreground of gravitational waves due to unresolved points sources~\cite{Farmer:2003pa,Rosado:2011kv,Adams:2013qma,TheLIGOScientific:2016wyq}. 

\subsection{Scaling relations}
\label{sec:scaling}
Note that the only mass scales which enter the generative sector potential are $\mu_{\phi}$ and $\Lambda_{\phi}$. Particularly useful scaling relations therefore hold when considering the parameter space of the model. A rescaling
	 \begin{equation}
	\mu_{\phi} \to \mu_{\phi}', \qquad \qquad \Lambda_{\phi}' \to \left(\frac{\mu_{\phi}'}{\mu_{\phi}} \right)\Lambda_{\phi},
	\label{eq:rescale}
	\end{equation}
leaves many important quantities invariant. In particular
	\begin{equation}
	\phi_{c}' =  \left( \frac{\mu_{\phi}'}{\mu_{\phi}}\right)\phi_{c}, \qquad  \qquad
	T_{c}' =  \left( \frac{\mu_{\phi}'}{\mu_{\phi}}\right)T_{c},
	\label{eq:rescaling1}
	\end{equation}
which leaves the ratio unchanged. Furthermore, when performing the rescaling in eq.~(\ref{eq:rescale}), the absolute value of the potential, e.g. at the barrier height, increases as $(\mu_{\phi}'/\mu_{\phi})^{4}$, but the bubble size decreases as $(\mu_{\phi}/\mu_{\phi}')$, which results in
	\begin{equation}
	\phi_{n}' =  \left( \frac{\mu_{\phi}'}{\mu_{\phi}}\right)\phi_{n}, \qquad  \qquad
	T_{n}' =  \left( \frac{\mu_{\phi}'}{\mu_{\phi}}\right)T_{n},
	\label{eq:rescaling2}
	\end{equation}
again leaving the ratio unchanged.\footnote{Up to a (negligible for our purposes) logarithmic correction to the relation in eq.~(\ref{eq:nucleation}), which comes from the temperature dependence of the Hubble rate.}
We also find
	\begin{align}
	\alpha' = \alpha, \qquad \qquad
	\left(\frac{\beta}{H}\right)' =  \left(\frac{\beta}{H}\right).
	\end{align}
This can be understood from eq.~(\ref{eq:rescaling2}). The increase in the magnitude of $\rho_{\rm vac}$ is cancelled by the increase in nucleation temperature in eq.~(\ref{eq:alpha}), leaving $\alpha$ unchanged.\footnote{Up to changes in $g_{\ast}(T)$ in the symmetric phase plasma, which we take to be constant for simplicity.} Similarly, the increase in $T_{n}$ is exactly cancelled by the smaller $S_{3}/T$ gradient in eq.~(\ref{eq:beta}), leaving $\beta/H$ unchanged. However, note the peak gravitational wave frequency, eq.~(\ref{eq:peakfreq}), does change, due to its linear dependence on $T_{n}$. These relations allow one to more efficiently scan the parameter space of the model, because the key quantities, $\alpha$, $\beta/H$ and $T_{n}$ can be determined in terms of a fixed values of $\mu_{\phi}$ and $\Lambda_{\phi}$ and then rescaled in a computationally efficient way.

\begin{figure}[t]
\begin{center}
\includegraphics[width=300pt]{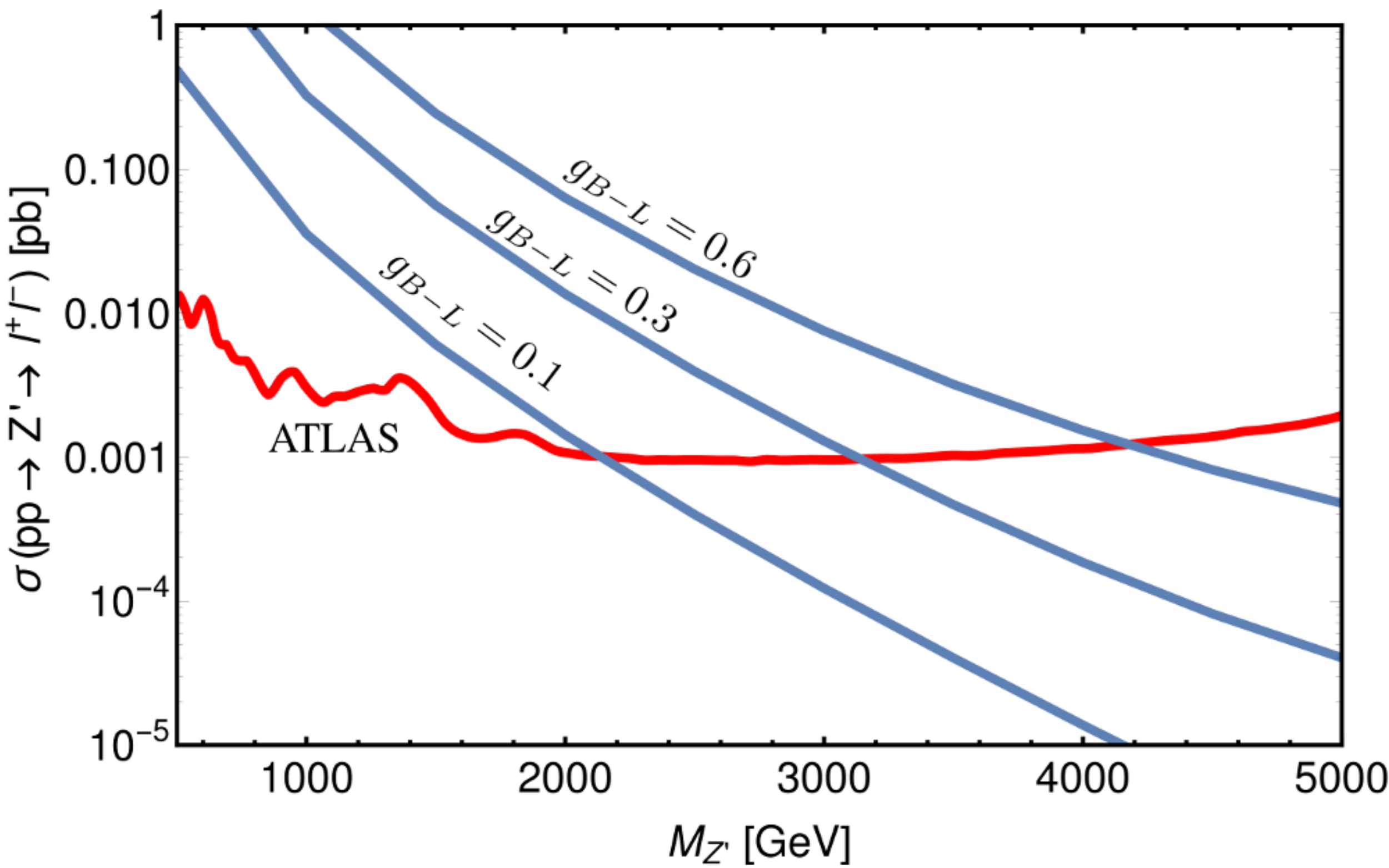}
\end{center}
\caption{Predicted resonant $Z'$ dilepton signal and limit from the ATLAS~\cite{ATLAS-CONF-2016-045} (CMS gives similar limits~\cite{Khachatryan:2016zqb}). The intrinsic width of the resonance is $\Gamma_{Z'}\approx 0.3 \%, \; 2.9 \%, \; 11 \%$ of $M_{Z'}$ for gauge couplings $g_{B-L}=0.1, \; 0.3, \; 0.6$ respectively.}
\label{fig:atlas13}
\end{figure}

\section{Additional signals}

\subsection{$Z_{B-L}'$ at the LHC}
The model introduces a $U(1)_{B-L}$ gauge group, which explains the presence of a global $B-L$ symmetry at low energy, crucial for DM-baryon symmetric models~\cite{Petraki:2011mv}. The associated $Z'$ can be searched for at the LHC~\cite{ATLAS-CONF-2016-045,Khachatryan:2016zqb}. The limit on narrow $Z'$ resonances decaying to $l^{+}l^{-}$, with $l=e,\mu$, from a search using 13 TeV ATLAS data is shown in figure~\ref{fig:atlas13}. Relevant formulas for determining the cross section for $Z'$ production can be found e.g. in~\cite{Clarke:2011aa}. We use the MSTW 2008 parton distribution functions~\cite{Martin:2009iq} and an estimate for the higher order QCD corrections $K=1.16$, as suggested by ATLAS for sequential standard model $Z'$ masses of 2 and 3 TeV. From figure~\ref{fig:atlas13} we see that the $B-L$ breaking scale should be $q_{B-L}^\sigma v_{B-L} \gtrsim \mathcal{O}(10)$ TeV, where $q_{B-L}^\sigma$ is the charge of the $U(1)_{B-L}$ breaking Higgs.    

\subsection{Halo ellipticity}
We now turn to the DM sector. The phenomenology of DM sectors with gauged $U(1)_{D}$ symmetries displays rich variations depending on the area of parameter space one is in. Here we will consider relatively large dark fine structure constants, $\alpha_{D} \equiv g_{D}^{2}/4\pi \gtrsim 0.1$, and DM constituent masses Min$[m_{\zeta}, m_{\xi}]\gtrsim 0.1$ GeV, for which the DM is sufficiently tightly bound to approach the collisionless limit. The binding energy is given by
	\begin{equation}
	\Delta \approx \frac{\alpha_{D}^{2}}{2} \mu \equiv \frac{\alpha_{D}^{2}}{2} \frac{ m_{\zeta} m_{\xi} }{ m_{\zeta}+ m_{\xi} }.
	\end{equation}
The most stringent constraints then come from halo ellipticity observations. For $M_{DM} = m_{\zeta}+  m_{\xi} = 1.5$ GeV there is significant tension with observations even with relatively large binding energies, $\Delta > 10$ MeV~\cite{CyrRacine:2012fz,Petraki:2014uza}. In this regime the ionized fraction of DM is completely negligible and the hyperfine transition energy is too large for dissipative cooling to be effective~\cite{CyrRacine:2012fz,Boddy:2016bbu}. If future numerical studies confirm the stringent halo ellipticity constraints such a scenario will be ruled out. The complicated phenomenology of dissipative regions of parameter space is beyond the scope of this work~\cite{Foot:2014uba}. The constraint can be evaded if the $U(1)_{D}$ symmetry is broken and we discuss this possibility further in section~\ref{sec:dd}.

\subsection{$\Delta N_{\rm eff}$}

We now turn to independent constraints on the model, coming from BBN and the CMB, which were derived in~\cite{Petraki:2011mv}. Since then, Planck has improved the limit on radiation at the CMB epoch, giving the effective number of neutrino species $N_{\rm eff}=3.15 \pm 0.23$~\cite{Ade:2015xua}. From this and the SM value $N_{\rm eff}=3.046$, we take $\Delta N_{\rm eff} < 0.6$ as a conservative (essentially $2\sigma$) limit on a BSM contribution. If $U(1)_{D}$ is unbroken, the dark photons will contribute to $\Delta N_{\rm eff}$ at the CMB epoch. The contribution depends on the degrees-of-freedom of the dark photon (two) and the dark radiation temperature. This allows one to translate the limit on extra radiation into a bound on the number of effective degrees of freedom of the dark sector, at the temperature it decouples from the visible sector radiation bath,
	\begin{equation}
	g_{\rm ds}(T_{\rm dec}) \lesssim 12.6 \left( \frac{\Delta N_{\rm eff}}{0.6}\right)^{3/4} \left( \frac{g_{\rm vs}(T_{\rm dec})}{110.25} \right),
	\end{equation}  
where $g_{\rm vs}$ are the effective visible sector degrees-of-freedom (110.25 when including the $f_{L,R}$ states). The dark sector has $g_{\rm ds}=16.5$ $(12.5)$ including (excluding) the contribution of $\chi$. From which we learn that $\chi$ should be heavy enough to not contribute to $g_{\rm ds}(T_{\rm dec})$ or additional degrees of freedom that contribute to $g_{\rm vs}$ must exist. If the couplings between the sectors are not too small, the former possibility will typically be the case, as the decoupling will be driven by the temperature falling sufficiently below the $\chi$ mass making the communication between the sectors (which proceeds through interactions with $\chi$) inefficient. Furthermore, future improvements on $\Delta N_{\rm eff}$, estimated to reach uncertainties of $\sigma(\Delta N_{\rm eff})=0.02$ for a stage-IV CMB experiment~\cite{Abazajian:2013oma}, should be able to rule out the unbroken $U(1)_{D}$ scenario independently of the halo ellipticity observations. In order for the above conclusions to hold, we require the kinetic mixing parameter between the SM $U(1)_{Y}$ and DM $U(1)_{D}$ field strength tensors, $\epsilon F_{\mu \nu} F^{\mu \nu}_{D}$, to be sufficiently small. Otherwise interactions between the lightest dark matter particle Min$[m_{\zeta},m_{\xi}]$ and the electrons can keep the sectors in equilibrium. Here, for notational clarity, we will assume $m_{\zeta}< m_{\xi}$ as the subsequent phenomenology does not depend on the mass ordering and we will also make the mild assumption $m_{e}<m_{\zeta}$. To keep the sectors unequilibriated, we then require~\cite{Petraki:2014uza}
	\begin{equation}
	\label{eq:epslimunbroken}
	\epsilon \lesssim \frac{ 10^{-6} }{ \alpha_{D} } \left( \frac{  m_{\zeta} }{ 0.1 \; \mathrm{GeV} } \right)^{3/2}.
	\end{equation}
This estimate is sufficiently accurate for our purposes and broadly consistent with bounds coming from detailed studies of the energy transfer between the visible and dark sectors in such models~\cite{Vogel:2013raa}.

\subsection{Direct detection}
\label{sec:dd}
We now update the direct detection constraints discussed in~\cite{Petraki:2011mv}. The spin-independent DM-nucleon scattering cross section via $Z'$ exchange is given by
	\begin{equation}
	\sigma_{B-L}^{\rm SI} \sim  10^{-45} \; \mathrm{cm}^{2} \left( \frac{g_{B-L}}{0.1} \right)^{4} \Big( \frac{ 2 \; \mathrm{TeV}}{ M_{Z'} } \Big)^{4}\left( \frac{\mu_{N}}{0.6 \; \mathrm{GeV}} \right)^{2},
	\end{equation}
where $\mu_{N}$ is the DM-Nucleon reduced mass. The current limit from CRESST-II, for $m_{DM}=1.5$ GeV, requires $\sigma^{\rm SI} \lesssim 2.7 \times 10^{-39} \; \mathrm{cm}^{2}$~\cite{Angloher:2015ewa}. However, taking into account the limit on $M_{Z'}$ from the LHC, this cross section is constrained to be below the floor for coherent neutrino scattering on CaWO$_{4}$, which for this $m_{DM}$, is $\sigma \approx 10^{-43} \; \mathrm{cm}^{2}$~\cite{Gutlein:2014gma}. As the DM forms tightly bound atomic states, the EM millicharge of the DM, induced by the kinetic mixing with the photon, is screened and the interaction becomes effectively contact type. The DM-proton scattering cross section becomes suppressed by the Bohr radius of the DM bound state to the fourth power, $1/(\alpha_{D}m_{\zeta})^{4}$, and can be approximated by~\cite{Cline:2012is}
	\begin{equation}
	\sigma_{\gamma}^{\rm SI} \sim 10^{-39} \; \mathrm{cm}^{2} \; \left( \frac{\epsilon}{10^{-6}} \right)^{2}\left( \frac{0.3}{\alpha_{D}} \right)^{3} \left( \frac{0.5 \; \mathrm{GeV}}{m_{\zeta}} \right)^{4}	 \left( \frac{\mu_{N}}{0.6 \; \mathrm{GeV}} \right)^{2}.
	\end{equation}
This can be above the neutrino floor. However, note that large signals, which depend on small values of $\alpha_{D}m_{\zeta}$, are in greater tension with the halo ellipticity constraints.

We now discuss the possibility of a broken $U(1)_{D}$ symmetry. For simplicity, here we consider parameter space in which the DM is no longer in a bound state but made up of unbound $\zeta$ and $\xi$ particles.\footnote{The intermediate regime in which the $U(1)_{D}$ is broken, but with a sufficiently light mediator for bound states to form was considered in~\cite{Petraki:2014uza}.
}
The DM with $\sim$ GeV scale mass can still annihilate efficiently into dark photons with mass $M_{D}\sim100$ MeV, which decay before BBN provided that $\epsilon \gtrsim 10^{-10}$. Beam dump and supernova constraints can be evaded provided that $M_{D} \gtrsim 200$ MeV (though some areas of parameter space below this are still allowed~\cite{Cirelli:2016rnw}). Limits from the CMB~\cite{Cirelli:2016rnw,Bringmann:2016din} do not apply if the DM antiparticle population is sufficiently depleted. The spin-independent DM-nucleon scattering cross section is then~\cite{Panci:2014gga}\footnote{This cross section applies only for $M_{D} \gtrsim 2E_{R}m_{T} \sim 1-10$ MeV, where $E_{R} \sim$ keV is the nuclear recoil and $m_{T} \sim 10-100$ GeV is the relevant target mass. For smaller $M_{D}$ the interaction becomes long range.}
	\begin{equation}
	\sigma_{D}^{\rm SI} \sim 10^{-40} \; \mathrm{cm}^{2} \; \left( \frac{\epsilon}{10^{-5}} \right)^{2} \left( \frac{\alpha_{D}}{10^{-2}} \right) \left( \frac{300 \; \mathrm{MeV} }{ M_{D} } \right)^{4}\left( \frac{\mu_{N}}{0.6 \; \mathrm{GeV}} \right)^{2}.
	\end{equation}
We have checked that this approximately reproduces the limits presented in~\cite{Cirelli:2016rnw}, in which a more careful treatment of the energy spectrum of nuclear recoils and the sensitivity of CRESST-II to low energy nuclear recoils is taken into account.

\section{Discussion and conclusion}
The detection of gravitational waves has opened up a new avenue for understanding the Universe. Future space based gravitational wave observatories will have sensitivity to first order phase transitions occurring at $T \sim \mathcal{O}$(TeV). One interesting possibility is a signal from a strong electroweak phase transition associated with electroweak baryogenesis~\cite{Grojean:2006bp,No:2011fi,Delaunay:2006ws,Vaskonen:2016yiu,Dorsch:2016nrg,Huang:2016odd}. Here we have considered a more exotic possibility, in which both dark matter and baryon asymmetries are generated during a strong phase transition associated with the spontaneous breaking of a yet undiscovered gauge symmetry.

Such a scenario necessitates a non-minimal structure, following the requirements for generating an asymmetry and transferring it to the visible and dark sectors. It is interesting to note that the large couplings --- which are required for CP violation and to strengthen the phase transition --- help to provide sufficient friction to keep the bubble wall in the non-runaway regime. This can allow for relatively strong phase transitions with subsonic walls which are both detectable by future gravitational wave observatories and self-consistent in the framework of asymmetry generation via an electroweak baryogenesis type mechanism. The possibility of gravitational wave signatures is expected to be a generic feature of such models, as they all require a strong first order phase transition associated with the breaking of an exotic non-abelian gauge symmetry, together with chiral BSM fermions which give the CP violation and help to provide friction~\cite{Dutta:2006pt,Shelton:2010ta,Dutta:2010va,Walker:2012ka,Davoudiasl:2016ijt}. The dark sector of this model also exhibits an interesting structure, which can contribute non-negligibly to $\Delta N_{\rm eff}$ or lead to DM-nucleon scattering. The case with an unbroken $U(1)_{D}$ gauge symmetry is in tension with halo ellipticity constraints on DM self interactions. The scenario evades EDM constraints as the CP violating interactions occur in the generative sector.

It is encouraging that upcoming experiments are sensitive to such exotic, non-minimal, scenarios. While the theory space is too large to pursue every possibility, it may be that non-null experimental results will in the near future help us uncover unexpected new physics. Of particular interest for the type of models considered here are space based gravitational wave observatories with sensitivity to frequencies in the range $\mathcal{O}(10^{-3}-10)$ Hz. The lower range will be probed by LISA and the higher range can be probed by follow-up-missions such as BBO and DECIGO~\cite{Crowder:2005nr,Kudoh:2005as,Kawamura:2006up}. The greater sensitivity expected of second generation space based interferometers will also greatly improve the prospects of detecting such phase transitions.

\subsection*{Acknowledgements}
I would like to thank Thomas Konstandin, Kalliopi Petraki and G{\'e}raldine Servant for useful discussion. I would also like to thank the organisers of the 2016 Gravitational Waves and Cosmology Workshop for an inspiring event.

\appendix

\section{Dimension four potential}
The simplest choice of tree level potential corresponds to a single field potential up to dimension four
	\begin{equation}
	V_{G} = \frac{ \mu_{\phi}^{2} }{ 2 }\phi^{2} + \frac{ \lambda_{ \phi }}{4} \phi^{4}.
	\label{eq:v4pot}
	\end{equation}
Such a potential can arise if the cross couplings of the generative scalar to the other scalars in the theory are negligible. The minimum of the potential is given by the usual expression $v_{\phi}=-\mu_{\phi}/\sqrt{ \lambda_{ \phi }}$. The relevant field dependent masses are those of the generative sector Higgs mass,
	\begin{equation}
	m_{\phi}^{2}  =\mu_{\phi}^{2}+3\lambda_{\phi}\phi^{2},
	\end{equation}
the $\Psi$ fermions due to the Yukawa interactions of eq.~(\ref{eq:genyuk}) and the $SU(2)_{G}$ gauge bosons.

Here the barrier in the effective potential can be created by the one loop thermal term for the gauge bosons. (If the gauge coupling is large enough a barrier can also be present at zero temperature.) The phase transition is strongly first order if the gauge coupling is large enough and the scalar quartic $\lambda_{\phi}$ is small enough~\cite{Carrington:1991hz}. The fermions help delay and hence strengthen the phase transition~\cite{Carena:2004ha}.

However, the large gauge couplings, $g_{G} \gtrsim 1$ for $\lambda_{\phi} \approx 0.05$, required in order to obtain a strong first order phase transition with the tree level potential in eq.~(\ref{eq:v4pot}), lead to pathological issues with the one-loop effective potential. The nucleation temperature becomes sensitive to the method used in implementing the daisy resummation. Quantitatively, we find large $\mathcal{O}(1)$ differences in the nucleation temperature when we include the daisy resummation as a separate term as opposed to when we include the thermal masses in the thermal functions directly~\cite{Curtin:2016urg} (see appendix~B). Hence the quantitative results for this potential are somewhat unreliable and we therefore study the $\phi^{6}$ potential for which these issues are not present because strong first order phase transitions with observable gravitational wave signals can then be achieved for smaller values of $g_{G}$ and larger values of $|\lambda_{\phi}|$.

\section{The effective potential}
For completeness we include here the form of the effective potential we have used in our calculations. We have used the one-loop effective potential in the on-shell renormalization scheme. The one-loop $T=0$ contribution is given by
	\begin{align}
	V_{1}^{0}(\phi) = \sum_{i} \frac{g_{i}(-1)^{F}}{64\pi^2} \bigg\{ m_{i}^{4}(\phi)\left(\mathrm{Ln}\left[\frac{m_{i}^{2}(\phi)}{m_{i}^{2}(v)} \right] - \frac{3}{2}\right)
			     +2m_{i}^{2}(\phi)m_{i}^{2}(v)\bigg\}, 	\label{eq:oneloop}
	\end{align}
where $g_{i}= \{1,9,16 \}$ for the $\phi$, $SU(2)_{G}$ gauge bosons and two generations of the $\Psi$ fermions, $F=0 \; (1)$ for bosons (fermions) and we have ignored the contribution of the Goldstone bosons. The one-loop finite $T$ contribution is given by
	\begin{equation}
	V_{1}^{T}(\phi,T)=\sum_{i} \frac{g_{i}(-1)^{F}T^{4}}{2\pi^2}\int_{0}^{\infty}y^{2} \; \mathrm{Ln}\left(1-(-1)^{F}\mathrm{Exp}\left[-\sqrt{ y^{2}+{m_{i}^{2}(\phi) }/{ T^{2} } }\right]\right) dy.
	\label{eq:finiteT}
	\end{equation}
The thermal masses for $\phi$ and the $SU(2)_{G}$ gauge bosons are given by
	\begin{align}
	\Pi_{\phi} &= \left(\frac{1}{2}\lambda_{\phi} + \frac{3}{16}g_{G}^2 + \frac{1}{6}h^2+\frac{1}{6}\tilde{h}^2 \right)T^{2}, \\
	\Pi_{G} &= \left( \frac{11}{6}g_{G}^2 \right)T^{2}.
	\end{align}
In order to take into account the resummation of the Matsubara zero modes one can include the daisy term
	\begin{equation}
	V_{\rm Daisy}(\phi,T) = \sum_{i}\frac{\overline{g}_{i}T}{12 \pi}\Big \{ m_{i}^{3}(\phi)-\big[m_{i}^{2}(\phi)+\Pi_{i}(T)\big]^{3/2}\Big\}
	\label{eq:daisy}
	\end{equation}
where the sum runs only over scalars and the longitudinal degrees of freedom of the vector bosons, i.e $\bar{g_{i}} \equiv \{ 1,3 \}$ for $\phi$ and the $SU(2)_{G}$ gauge bosons. The effective potential is then given by the sum of the tree level potential and Eqs.~(\ref{eq:oneloop}), (\ref{eq:finiteT}) and (\ref{eq:daisy}). An alternative prescription is to substitute $m_{i}^{2} \to m_{i}^{2} +  \Pi_{i}$ in eq.~(\ref{eq:finiteT}) for scalars and longitudinal gauge bosons~\cite{Curtin:2016urg}. The latter form has the advantage in that it correctly captures the decoupling of heavy degrees of freedom from the effective potential. However, it has the disadvantage in that it effectively resums all Matsubara modes, not just the zero modes. Hence for the $\phi^{4}$ potential, where the two alternative approaches return significantly different results (see appendix~A), we can no longer necessarily trust the quantitative predictions coming from the effective potential.

\bibliographystyle{JHEP}   

\begin{thebibliography}{100}

\bibitem{Nussinov:1985xr}
S.~Nussinov, \emph{{Technocosmology: could a technibaryon excess provide a
  `natural' missing mass candidate?}},
  \href{http://dx.doi.org/10.1016/0370-2693(85)90689-6}{\emph{Phys. Lett.} {\bf
  B165} (1985) 55--58}.

\bibitem{Davoudiasl:2012uw}
H.~Davoudiasl and R.~N. Mohapatra, \emph{{On Relating the Genesis of Cosmic
  Baryons and Dark Matter}},
  \href{http://dx.doi.org/10.1088/1367-2630/14/9/095011}{\emph{New J. Phys.}
  {\bf 14} (2012) 095011}, [\href{http://arxiv.org/abs/1203.1247}{{\tt
  1203.1247}}].

\bibitem{Petraki:2013wwa}
K.~Petraki and R.~R. Volkas, \emph{{Review of asymmetric dark matter}},
  \href{http://dx.doi.org/10.1142/S0217751X13300287}{\emph{Int. J. Mod. Phys.}
  {\bf A28} (2013) 1330028}, [\href{http://arxiv.org/abs/1305.4939}{{\tt
  1305.4939}}].

\bibitem{Zurek:2013wia}
K.~M. Zurek, \emph{{Asymmetric Dark Matter: Theories, Signatures, and
  Constraints}},
  \href{http://dx.doi.org/10.1016/j.physrep.2013.12.001}{\emph{Phys. Rept.}
  {\bf 537} (2014) 91--121}, [\href{http://arxiv.org/abs/1308.0338}{{\tt
  1308.0338}}].

\bibitem{Sakharov:1967dj}
A.~D. Sakharov, \emph{{Violation of CP Invariance, c Asymmetry, and Baryon
  Asymmetry of the Universe}},
  \href{http://dx.doi.org/10.1070/PU1991v034n05ABEH002497}{\emph{Pisma Zh.
  Eksp. Teor. Fiz.} {\bf 5} (1967) 32--35}.

\bibitem{Chiu:1966kg}
H.-Y. Chiu, \emph{{Symmetry between particle and anti-particle populations in
  the universe}},
  \href{http://dx.doi.org/10.1103/PhysRevLett.17.712}{\emph{Phys. Rev. Lett.}
  {\bf 17} (1966) 712}.

\bibitem{Petraki:2011mv}
K.~Petraki, M.~Trodden and R.~R. Volkas, \emph{{Visible and dark matter from a
  first-order phase transition in a baryon-symmetric universe}},
  \href{http://dx.doi.org/10.1088/1475-7516/2012/02/044}{\emph{JCAP} {\bf 1202}
  (2012) 044}, [\href{http://arxiv.org/abs/1111.4786}{{\tt 1111.4786}}].

\bibitem{Kuzmin:1985mm}
V.~A. Kuzmin, V.~A. Rubakov and M.~E. Shaposhnikov, \emph{{On the Anomalous
  Electroweak Baryon Number Nonconservation in the Early Universe}},
  \href{http://dx.doi.org/10.1016/0370-2693(85)91028-7}{\emph{Phys. Lett.} {\bf
  B155} (1985) 36}.

\bibitem{Shaposhnikov:1986jp}
M.~E. Shaposhnikov, \emph{{Possible Appearance of the Baryon Asymmetry of the
  Universe in an Electroweak Theory}}, {\emph{JETP Lett.} {\bf 44} (1986)
  465--468}.

\bibitem{Shaposhnikov:1987tw}
M.~E. Shaposhnikov, \emph{{Baryon Asymmetry of the Universe in Standard
  Electroweak Theory}},
  \href{http://dx.doi.org/10.1016/0550-3213(87)90127-1}{\emph{Nucl. Phys.} {\bf
  B287} (1987) 757--775}.

\bibitem{Cohen:1993nk}
A.~G. Cohen, D.~B. Kaplan and A.~E. Nelson, \emph{{Progress in electroweak
  baryogenesis}},
  \href{http://dx.doi.org/10.1146/annurev.ns.43.120193.000331}{\emph{Ann. Rev.
  Nucl. Part. Sci.} {\bf 43} (1993) 27--70},
  [\href{http://arxiv.org/abs/hep-ph/9302210}{{\tt hep-ph/9302210}}].

\bibitem{Trodden:1998ym}
M.~Trodden, \emph{{Electroweak baryogenesis}},
  \href{http://dx.doi.org/10.1103/RevModPhys.71.1463}{\emph{Rev. Mod. Phys.}
  {\bf 71} (1999) 1463--1500}, [\href{http://arxiv.org/abs/hep-ph/9803479}{{\tt
  hep-ph/9803479}}].

\bibitem{Morrissey:2012db}
D.~E. Morrissey and M.~J. Ramsey-Musolf, \emph{{Electroweak baryogenesis}},
  \href{http://dx.doi.org/10.1088/1367-2630/14/12/125003}{\emph{New J. Phys.}
  {\bf 14} (2012) 125003}, [\href{http://arxiv.org/abs/1206.2942}{{\tt
  1206.2942}}].

\bibitem{Dutta:2006pt}
B.~Dutta and J.~Kumar, \emph{{Hidden sector baryogenesis}},
  \href{http://dx.doi.org/10.1016/j.physletb.2006.09.069}{\emph{Phys. Lett.}
  {\bf B643} (2006) 284--289}, [\href{http://arxiv.org/abs/hep-th/0608188}{{\tt
  hep-th/0608188}}].

\bibitem{Shelton:2010ta}
J.~Shelton and K.~M. Zurek, \emph{{Darkogenesis: A baryon asymmetry from the
  dark matter sector}},
  \href{http://dx.doi.org/10.1103/PhysRevD.82.123512}{\emph{Phys. Rev.} {\bf
  D82} (2010) 123512}, [\href{http://arxiv.org/abs/1008.1997}{{\tt
  1008.1997}}].

\bibitem{Dutta:2010va}
B.~Dutta and J.~Kumar, \emph{{Asymmetric Dark Matter from Hidden Sector
  Baryogenesis}},
  \href{http://dx.doi.org/10.1016/j.physletb.2011.04.036}{\emph{Phys. Lett.}
  {\bf B699} (2011) 364--367}, [\href{http://arxiv.org/abs/1012.1341}{{\tt
  1012.1341}}].

\bibitem{Walker:2012ka}
D.~G.~E. Walker, \emph{{Dark Baryogenesis}},
  \href{http://arxiv.org/abs/1202.2348}{{\tt 1202.2348}}.

\bibitem{Davoudiasl:2016ijt}
H.~Davoudiasl, P.~P. Giardino and C.~Zhang, \emph{{Muon $g-2$ from Asymmetric
  Dark Matter in the Exo-Higgs Scenario}},
  \href{http://arxiv.org/abs/1612.05639}{{\tt 1612.05639}}.

\bibitem{Chatrchyan:2012xdj}
{\scshape CMS} collaboration, S.~Chatrchyan et~al., \emph{{Observation of a new
  boson at a mass of 125 GeV with the CMS experiment at the LHC}},
  \href{http://dx.doi.org/10.1016/j.physletb.2012.08.021}{\emph{Phys. Lett.}
  {\bf B716} (2012) 30--61}, [\href{http://arxiv.org/abs/1207.7235}{{\tt
  1207.7235}}].

\bibitem{Aad:2012tfa}
{\scshape ATLAS} collaboration, G.~Aad et~al., \emph{{Observation of a new
  particle in the search for the Standard Model Higgs boson with the ATLAS
  detector at the LHC}},
  \href{http://dx.doi.org/10.1016/j.physletb.2012.08.020}{\emph{Phys. Lett.}
  {\bf B716} (2012) 1--29}, [\href{http://arxiv.org/abs/1207.7214}{{\tt
  1207.7214}}].

\bibitem{ATLAS-CONF-2016-045}
{\scshape ATLAS} collaboration, \emph{{Search for new high-mass resonances in
  the dilepton final state using proton-proton collisions at $\sqrt{s}$ = 13
  TeV with the ATLAS detector}},  Tech. Rep. ATLAS-CONF-2016-045, CERN, Geneva,
  Aug, 2016.

\bibitem{Khachatryan:2016zqb}
{\scshape CMS} collaboration, V.~Khachatryan et~al., \emph{{Search for narrow
  resonances in dilepton mass spectra in proton-proton collisions at $\sqrt{s}$
  = 13 TeV and combination with 8 TeV data}},
  \href{http://arxiv.org/abs/1609.05391}{{\tt 1609.05391}}.

\bibitem{Ade:2015xua}
{\scshape Planck} collaboration, P.~A.~R. Ade et~al., \emph{{Planck 2015
  results. XIII. Cosmological parameters}},
  \href{http://dx.doi.org/10.1051/0004-6361/201525830}{\emph{Astron.
  Astrophys.} {\bf 594} (2016) A13},
  [\href{http://arxiv.org/abs/1502.01589}{{\tt 1502.01589}}].

\bibitem{Abbott:2016blz}
{\scshape Virgo, LIGO Scientific} collaboration, B.~P. Abbott et~al.,
  \emph{{Observation of Gravitational Waves from a Binary Black Hole Merger}},
  \href{http://dx.doi.org/10.1103/PhysRevLett.116.061102}{\emph{Phys. Rev.
  Lett.} {\bf 116} (2016) 061102}, [\href{http://arxiv.org/abs/1602.03837}{{\tt
  1602.03837}}].

\bibitem{Abbott:2016nmj}
{\scshape Virgo, LIGO Scientific} collaboration, B.~P. Abbott et~al.,
  \emph{{GW151226: Observation of Gravitational Waves from a 22-Solar-Mass
  Binary Black Hole Coalescence}},
  \href{http://dx.doi.org/10.1103/PhysRevLett.116.241103}{\emph{Phys. Rev.
  Lett.} {\bf 116} (2016) 241103}, [\href{http://arxiv.org/abs/1606.04855}{{\tt
  1606.04855}}].

\bibitem{Armano:2016bkm}
M.~Armano et~al., \emph{{Sub-Femto-g Free Fall for Space-Based Gravitational
  Wave Observatories: LISA Pathfinder Results}},
  \href{http://dx.doi.org/10.1103/PhysRevLett.116.231101}{\emph{Phys. Rev.
  Lett.} {\bf 116} (2016) 231101}.

\bibitem{Hindmarsh:2013xza}
M.~Hindmarsh, S.~J. Huber, K.~Rummukainen and D.~J. Weir, \emph{{Gravitational
  waves from the sound of a first order phase transition}},
  \href{http://dx.doi.org/10.1103/PhysRevLett.112.041301}{\emph{Phys. Rev.
  Lett.} {\bf 112} (2014) 041301}, [\href{http://arxiv.org/abs/1304.2433}{{\tt
  1304.2433}}].

\bibitem{Hindmarsh:2015qta}
M.~Hindmarsh, S.~J. Huber, K.~Rummukainen and D.~J. Weir, \emph{{Numerical
  simulations of acoustically generated gravitational waves at a first order
  phase transition}},
  \href{http://dx.doi.org/10.1103/PhysRevD.92.123009}{\emph{Phys. Rev.} {\bf
  D92} (2015) 123009}, [\href{http://arxiv.org/abs/1504.03291}{{\tt
  1504.03291}}].

\bibitem{Hindmarsh:2016lnk}
M.~Hindmarsh, \emph{{Sound shell model for acoustic gravitational wave
  production at a first-order phase transition in the early Universe}},
  \href{http://arxiv.org/abs/1608.04735}{{\tt 1608.04735}}.

\bibitem{Grojean:2006bp}
C.~Grojean and G.~Servant, \emph{{Gravitational Waves from Phase Transitions at
  the Electroweak Scale and Beyond}},
  \href{http://dx.doi.org/10.1103/PhysRevD.75.043507}{\emph{Phys. Rev.} {\bf
  D75} (2007) 043507}, [\href{http://arxiv.org/abs/hep-ph/0607107}{{\tt
  hep-ph/0607107}}].

\bibitem{Buchmuller:2013lra}
W.~Buchm{\"u}ller, V.~Domcke, K.~Kamada and K.~Schmitz, \emph{{The
  Gravitational Wave Spectrum from Cosmological $B-L$ Breaking}},
  \href{http://dx.doi.org/10.1088/1475-7516/2013/10/003}{\emph{JCAP} {\bf 1310}
  (2013) 003}, [\href{http://arxiv.org/abs/1305.3392}{{\tt 1305.3392}}].

\bibitem{Schwaller:2015tja}
P.~Schwaller, \emph{{Gravitational Waves from a Dark Phase Transition}},
  \href{http://dx.doi.org/10.1103/PhysRevLett.115.181101}{\emph{Phys. Rev.
  Lett.} {\bf 115} (2015) 181101}, [\href{http://arxiv.org/abs/1504.07263}{{\tt
  1504.07263}}].

\bibitem{Jaeckel:2016jlh}
J.~Jaeckel, V.~V. Khoze and M.~Spannowsky, \emph{{Hearing the signals of dark
  sectors with gravitational wave detectors}},
  \href{http://dx.doi.org/10.1103/PhysRevD.94.103519}{\emph{Phys. Rev.} {\bf
  D94} (2016) 103519}, [\href{http://arxiv.org/abs/1602.03901}{{\tt
  1602.03901}}].

\bibitem{Dev:2016feu}
P.~S.~B. Dev and A.~Mazumdar, \emph{{Probing the Scale of New Physics by
  Advanced LIGO/VIRGO}},
  \href{http://dx.doi.org/10.1103/PhysRevD.93.104001}{\emph{Phys. Rev.} {\bf
  D93} (2016) 104001}, [\href{http://arxiv.org/abs/1602.04203}{{\tt
  1602.04203}}].

\bibitem{Davoudiasl:2016yfa}
H.~Davoudiasl, P.~P. Giardino and C.~Zhang, \emph{{Higgs-like boson at 750 GeV
  and genesis of baryons}},
  \href{http://dx.doi.org/10.1103/PhysRevD.94.015006}{\emph{Phys. Rev.} {\bf
  D94} (2016) 015006}, [\href{http://arxiv.org/abs/1605.00037}{{\tt
  1605.00037}}].

\bibitem{Kobakhidze:2016mch}
A.~Kobakhidze, A.~Manning and J.~Yue, \emph{{Gravitational Waves from the Phase
  Transition of a Non-linearly Realised Electroweak Gauge Symmetry}},
  \href{http://arxiv.org/abs/1607.00883}{{\tt 1607.00883}}.

\bibitem{GarciaGarcia:2016xgv}
I.~Garcia~Garcia, S.~Krippendorf and J.~March-Russell, \emph{{The String
  Soundscape at Gravitational Wave Detectors}},
  \href{http://arxiv.org/abs/1607.06813}{{\tt 1607.06813}}.

\bibitem{Addazi:2016fbj}
A.~Addazi, \emph{{Limiting First Order Phase Transitions in Dark Gauge Sectors
  from Gravitational Waves experiments}},
  \href{http://arxiv.org/abs/1607.08057}{{\tt 1607.08057}}.

\bibitem{Katz:2016adq}
A.~Katz and A.~Riotto, \emph{{Baryogenesis and Gravitational Waves from Runaway
  Bubble Collisions}},
  \href{http://dx.doi.org/10.1088/1475-7516/2016/11/011}{\emph{JCAP} {\bf 1611}
  (2016) 011}, [\href{http://arxiv.org/abs/1608.00583}{{\tt 1608.00583}}].

\bibitem{Balazs:2016tbi}
C.~Balazs, A.~Fowlie, A.~Mazumdar and G.~White, \emph{{Gravitational waves at
  aLIGO and vacuum stability with a scalar singlet extension of the Standard
  Model}}, \href{http://dx.doi.org/10.1103/PhysRevD.95.043505}{\emph{Phys.
  Rev.} {\bf D95} (2017) 043505}, [\href{http://arxiv.org/abs/1611.01617}{{\tt
  1611.01617}}].

\bibitem{Huang:2017laj}
F.~P. Huang and X.~Zhang, \emph{{Probing the hidden gauge symmetry breaking
  through the phase transition gravitational waves}},
  \href{http://arxiv.org/abs/1701.04338}{{\tt 1701.04338}}.

\bibitem{Baker:2006ts}
C.~A. Baker et~al., \emph{{An Improved experimental limit on the electric
  dipole moment of the neutron}},
  \href{http://dx.doi.org/10.1103/PhysRevLett.97.131801}{\emph{Phys. Rev.
  Lett.} {\bf 97} (2006) 131801},
  [\href{http://arxiv.org/abs/hep-ex/0602020}{{\tt hep-ex/0602020}}].

\bibitem{Hudson:2011zz}
J.~J. Hudson, D.~M. Kara, I.~J. Smallman, B.~E. Sauer, M.~R. Tarbutt and E.~A.
  Hinds, \emph{{Improved measurement of the shape of the electron}},
  \href{http://dx.doi.org/10.1038/nature10104}{\emph{Nature} {\bf 473} (2011)
  493--496}.

\bibitem{Baron:2013eja}
{\scshape ACME} collaboration, J.~Baron et~al., \emph{{Order of Magnitude
  Smaller Limit on the Electric Dipole Moment of the Electron}},
  \href{http://dx.doi.org/10.1126/science.1248213}{\emph{Science} {\bf 343}
  (2014) 269--272}, [\href{http://arxiv.org/abs/1310.7534}{{\tt 1310.7534}}].

\bibitem{Chao:2014dpa}
W.~Chao and M.~J. Ramsey-Musolf, \emph{{Electroweak Baryogenesis, Electric
  Dipole Moments, and Higgs Diphoton Decays}},
  \href{http://dx.doi.org/10.1007/JHEP10(2014)180}{\emph{JHEP} {\bf 10} (2014)
  180}, [\href{http://arxiv.org/abs/1406.0517}{{\tt 1406.0517}}].

\bibitem{Dorsch:2016nrg}
G.~C. Dorsch, S.~J. Huber, T.~Konstandin and J.~M. No, \emph{{A Second Higgs
  Doublet in the Early Universe: Baryogenesis and Gravitational Waves}},
  \href{http://arxiv.org/abs/1611.05874}{{\tt 1611.05874}}.

\bibitem{Balazs:2016yvi}
C.~Balazs, G.~White and J.~Yue, \emph{{Effective field theory, electric dipole
  moments and electroweak baryogenesis}},
  \href{http://arxiv.org/abs/1612.01270}{{\tt 1612.01270}}.

\bibitem{Witten:1982fp}
E.~Witten, \emph{{An SU(2) Anomaly}},
  \href{http://dx.doi.org/10.1016/0370-2693(82)90728-6}{\emph{Phys. Lett.} {\bf
  B117} (1982) 324--328}.

\bibitem{Grojean:2004xa}
C.~Grojean, G.~Servant and J.~D. Wells, \emph{{First-order electroweak phase
  transition in the standard model with a low cutoff}},
  \href{http://dx.doi.org/10.1103/PhysRevD.71.036001}{\emph{Phys. Rev.} {\bf
  D71} (2005) 036001}, [\href{http://arxiv.org/abs/hep-ph/0407019}{{\tt
  hep-ph/0407019}}].

\bibitem{Bodeker:2004ws}
D.~Bodeker, L.~Fromme, S.~J. Huber and M.~Seniuch, \emph{{The Baryon asymmetry
  in the standard model with a low cut-off}},
  \href{http://dx.doi.org/10.1088/1126-6708/2005/02/026}{\emph{JHEP} {\bf 02}
  (2005) 026}, [\href{http://arxiv.org/abs/hep-ph/0412366}{{\tt
  hep-ph/0412366}}].

\bibitem{Delaunay:2007wb}
C.~Delaunay, C.~Grojean and J.~D. Wells, \emph{{Dynamics of Non-renormalizable
  Electroweak Symmetry Breaking}},
  \href{http://dx.doi.org/10.1088/1126-6708/2008/04/029}{\emph{JHEP} {\bf 04}
  (2008) 029}, [\href{http://arxiv.org/abs/0711.2511}{{\tt 0711.2511}}].

\bibitem{Damgaard:2015con}
P.~H. Damgaard, A.~Haarr, D.~O'Connell and A.~Tranberg, \emph{{Effective Field
  Theory and Electroweak Baryogenesis in the Singlet-Extended Standard Model}},
  \href{http://dx.doi.org/10.1007/JHEP02(2016)107}{\emph{JHEP} {\bf 02} (2016)
  107}, [\href{http://arxiv.org/abs/1512.01963}{{\tt 1512.01963}}].

\bibitem{Quiros:1999jp}
M.~Quiros, \emph{{Finite temperature field theory and phase transitions}},  in
  \emph{{Proceedings, Summer School in High-energy physics and cosmology:
  Trieste, Italy, June 29-July 17, 1998}}, pp.~187--259, 1999.
\newblock \href{http://arxiv.org/abs/hep-ph/9901312}{{\tt hep-ph/9901312}}.

\bibitem{Carson:1990jm}
L.~Carson, X.~Li, L.~D. McLerran and R.-T. Wang, \emph{{Exact Computation of
  the Small Fluctuation Determinant Around a Sphaleron}},
  \href{http://dx.doi.org/10.1103/PhysRevD.42.2127}{\emph{Phys. Rev.} {\bf D42}
  (1990) 2127--2143}.

\bibitem{Klinkhamer:1984di}
F.~R. Klinkhamer and N.~S. Manton, \emph{{A Saddle Point Solution in the
  Weinberg-Salam Theory}},
  \href{http://dx.doi.org/10.1103/PhysRevD.30.2212}{\emph{Phys. Rev.} {\bf D30}
  (1984) 2212}.

\bibitem{Linde:1980tt}
A.~D. Linde, \emph{{Fate of the False Vacuum at Finite Temperature: Theory and
  Applications}},
  \href{http://dx.doi.org/10.1016/0370-2693(81)90281-1}{\emph{Phys. Lett.} {\bf
  B100} (1981) 37--40}.

\bibitem{Anderson:1991zb}
G.~W. Anderson and L.~J. Hall, \emph{{The Electroweak phase transition and
  baryogenesis}}, \href{http://dx.doi.org/10.1103/PhysRevD.45.2685}{\emph{Phys.
  Rev.} {\bf D45} (1992) 2685--2698}.

\bibitem{Espinosa:2010hh}
J.~R. Espinosa, T.~Konstandin, J.~M. No and G.~Servant, \emph{{Energy Budget of
  Cosmological First-order Phase Transitions}},
  \href{http://dx.doi.org/10.1088/1475-7516/2010/06/028}{\emph{JCAP} {\bf 1006}
  (2010) 028}, [\href{http://arxiv.org/abs/1004.4187}{{\tt 1004.4187}}].

\bibitem{Caprini:2015zlo}
C.~Caprini et~al., \emph{{Science with the space-based interferometer eLISA.
  II: Gravitational waves from cosmological phase transitions}},
  \href{http://dx.doi.org/10.1088/1475-7516/2016/04/001}{\emph{JCAP} {\bf 1604}
  (2016) 001}, [\href{http://arxiv.org/abs/1512.06239}{{\tt 1512.06239}}].

\bibitem{Thrane:2013oya}
E.~Thrane and J.~D. Romano, \emph{{Sensitivity curves for searches for
  gravitational-wave backgrounds}},
  \href{http://dx.doi.org/10.1103/PhysRevD.88.124032}{\emph{Phys. Rev.} {\bf
  D88} (2013) 124032}, [\href{http://arxiv.org/abs/1310.5300}{{\tt
  1310.5300}}].

\bibitem{Moore:1995si}
G.~D. Moore and T.~Prokopec, \emph{{How fast can the wall move? A Study of the
  electroweak phase transition dynamics}},
  \href{http://dx.doi.org/10.1103/PhysRevD.52.7182}{\emph{Phys. Rev.} {\bf D52}
  (1995) 7182--7204}, [\href{http://arxiv.org/abs/hep-ph/9506475}{{\tt
  hep-ph/9506475}}].

\bibitem{Moore:2000wx}
G.~D. Moore, \emph{{Electroweak bubble wall friction: Analytic results}},
  \href{http://dx.doi.org/10.1088/1126-6708/2000/03/006}{\emph{JHEP} {\bf 03}
  (2000) 006}, [\href{http://arxiv.org/abs/hep-ph/0001274}{{\tt
  hep-ph/0001274}}].

\bibitem{Megevand:2009gh}
A.~Megevand and A.~D. Sanchez, \emph{{Velocity of electroweak bubble walls}},
  \href{http://dx.doi.org/10.1016/j.nuclphysb.2009.09.019}{\emph{Nucl. Phys.}
  {\bf B825} (2010) 151--176}, [\href{http://arxiv.org/abs/0908.3663}{{\tt
  0908.3663}}].

\bibitem{Konstandin:2010dm}
T.~Konstandin and J.~M. No, \emph{{Hydrodynamic obstruction to bubble
  expansion}},
  \href{http://dx.doi.org/10.1088/1475-7516/2011/02/008}{\emph{JCAP} {\bf 1102}
  (2011) 008}, [\href{http://arxiv.org/abs/1011.3735}{{\tt 1011.3735}}].

\bibitem{Huber:2011aa}
S.~J. Huber and M.~Sopena, \emph{{The bubble wall velocity in the minimal
  supersymmetric light stop scenario}},
  \href{http://dx.doi.org/10.1103/PhysRevD.85.103507}{\emph{Phys. Rev.} {\bf
  D85} (2012) 103507}, [\href{http://arxiv.org/abs/1112.1888}{{\tt
  1112.1888}}].

\bibitem{Huber:2013kj}
S.~J. Huber and M.~Sopena, \emph{{An efficient approach to electroweak bubble
  velocities}},  \href{http://arxiv.org/abs/1302.1044}{{\tt 1302.1044}}.

\bibitem{Megevand:2013hwa}
A.~M{\'e}gevand, \emph{{Friction forces on phase transition fronts}},
  \href{http://dx.doi.org/10.1088/1475-7516/2013/07/045}{\emph{JCAP} {\bf 1307}
  (2013) 045}, [\href{http://arxiv.org/abs/1303.4233}{{\tt 1303.4233}}].

\bibitem{Kozaczuk:2014kva}
J.~Kozaczuk, S.~Profumo, L.~S. Haskins and C.~L. Wainwright,
  \emph{{Cosmological Phase Transitions and their Properties in the NMSSM}},
  \href{http://dx.doi.org/10.1007/JHEP01(2015)144}{\emph{JHEP} {\bf 01} (2015)
  144}, [\href{http://arxiv.org/abs/1407.4134}{{\tt 1407.4134}}].

\bibitem{Kozaczuk:2015owa}
J.~Kozaczuk, \emph{{Bubble Expansion and the Viability of Singlet-Driven
  Electroweak Baryogenesis}},
  \href{http://dx.doi.org/10.1007/JHEP10(2015)135}{\emph{JHEP} {\bf 10} (2015)
  135}, [\href{http://arxiv.org/abs/1506.04741}{{\tt 1506.04741}}].

\bibitem{No:2011fi}
J.~M. No, \emph{{Large Gravitational Wave Background Signals in Electroweak
  Baryogenesis Scenarios}},
  \href{http://dx.doi.org/10.1103/PhysRevD.84.124025}{\emph{Phys. Rev.} {\bf
  D84} (2011) 124025}, [\href{http://arxiv.org/abs/1103.2159}{{\tt
  1103.2159}}].

\bibitem{Bodeker:2009qy}
D.~Bodeker and G.~D. Moore, \emph{{Can electroweak bubble walls run away?}},
  \href{http://dx.doi.org/10.1088/1475-7516/2009/05/009}{\emph{JCAP} {\bf 0905}
  (2009) 009}, [\href{http://arxiv.org/abs/0903.4099}{{\tt 0903.4099}}].

\bibitem{Creminelli:2001th}
P.~Creminelli, A.~Nicolis and R.~Rattazzi, \emph{{Holography and the
  electroweak phase transition}},
  \href{http://dx.doi.org/10.1088/1126-6708/2002/03/051}{\emph{JHEP} {\bf 03}
  (2002) 051}, [\href{http://arxiv.org/abs/hep-th/0107141}{{\tt
  hep-th/0107141}}].

\bibitem{Nardini:2007me}
G.~Nardini, M.~Quiros and A.~Wulzer, \emph{{A Confining Strong First-Order
  Electroweak Phase Transition}},
  \href{http://dx.doi.org/10.1088/1126-6708/2007/09/077}{\emph{JHEP} {\bf 09}
  (2007) 077}, [\href{http://arxiv.org/abs/0706.3388}{{\tt 0706.3388}}].

\bibitem{Randall:2006py}
L.~Randall and G.~Servant, \emph{{Gravitational waves from warped spacetime}},
  \href{http://dx.doi.org/10.1088/1126-6708/2007/05/054}{\emph{JHEP} {\bf 05}
  (2007) 054}, [\href{http://arxiv.org/abs/hep-ph/0607158}{{\tt
  hep-ph/0607158}}].

\bibitem{Konstandin:2010cd}
T.~Konstandin, G.~Nardini and M.~Quiros, \emph{{Gravitational Backreaction
  Effects on the Holographic Phase Transition}},
  \href{http://dx.doi.org/10.1103/PhysRevD.82.083513}{\emph{Phys. Rev.} {\bf
  D82} (2010) 083513}, [\href{http://arxiv.org/abs/1007.1468}{{\tt
  1007.1468}}].

\bibitem{Konstandin:2011dr}
T.~Konstandin and G.~Servant, \emph{{Cosmological Consequences of Nearly
  Conformal Dynamics at the TeV scale}},
  \href{http://dx.doi.org/10.1088/1475-7516/2011/12/009}{\emph{JCAP} {\bf 1112}
  (2011) 009}, [\href{http://arxiv.org/abs/1104.4791}{{\tt 1104.4791}}].

\bibitem{Konstandin:2011ds}
T.~Konstandin and G.~Servant, \emph{{Natural Cold Baryogenesis from Strongly
  Interacting Electroweak Symmetry Breaking}},
  \href{http://dx.doi.org/10.1088/1475-7516/2011/07/024}{\emph{JCAP} {\bf 1107}
  (2011) 024}, [\href{http://arxiv.org/abs/1104.4793}{{\tt 1104.4793}}].

\bibitem{Farmer:2003pa}
A.~J. Farmer and E.~S. Phinney, \emph{{The gravitational wave background from
  cosmological compact binaries}},
  \href{http://dx.doi.org/10.1111/j.1365-2966.2003.07176.x}{\emph{Mon. Not.
  Roy. Astron. Soc.} {\bf 346} (2003) 1197},
  [\href{http://arxiv.org/abs/astro-ph/0304393}{{\tt astro-ph/0304393}}].

\bibitem{Rosado:2011kv}
P.~A. Rosado, \emph{{Gravitational wave background from binary systems}},
  \href{http://dx.doi.org/10.1103/PhysRevD.84.084004}{\emph{Phys. Rev.} {\bf
  D84} (2011) 084004}, [\href{http://arxiv.org/abs/1106.5795}{{\tt
  1106.5795}}].

\bibitem{Adams:2013qma}
M.~R. Adams and N.~J. Cornish, \emph{{Detecting a Stochastic Gravitational Wave
  Background in the presence of a Galactic Foreground and Instrument Noise}},
  \href{http://dx.doi.org/10.1103/PhysRevD.89.022001}{\emph{Phys. Rev.} {\bf
  D89} (2014) 022001}, [\href{http://arxiv.org/abs/1307.4116}{{\tt
  1307.4116}}].

\bibitem{TheLIGOScientific:2016wyq}
{\scshape Virgo, LIGO Scientific} collaboration, B.~P. Abbott et~al.,
  \emph{{GW150914: Implications for the stochastic gravitational wave
  background from binary black holes}},
  \href{http://dx.doi.org/10.1103/PhysRevLett.116.131102}{\emph{Phys. Rev.
  Lett.} {\bf 116} (2016) 131102}, [\href{http://arxiv.org/abs/1602.03847}{{\tt
  1602.03847}}].

\bibitem{Clarke:2011aa}
J.~D. Clarke, R.~Foot and R.~R. Volkas, \emph{{Quark-lepton symmetric model at
  the LHC}}, \href{http://dx.doi.org/10.1103/PhysRevD.85.074012}{\emph{Phys.
  Rev.} {\bf D85} (2012) 074012}, [\href{http://arxiv.org/abs/1112.3405}{{\tt
  1112.3405}}].

\bibitem{Martin:2009iq}
A.~D. Martin, W.~J. Stirling, R.~S. Thorne and G.~Watt, \emph{{Parton
  distributions for the LHC}},
  \href{http://dx.doi.org/10.1140/epjc/s10052-009-1072-5}{\emph{Eur. Phys. J.}
  {\bf C63} (2009) 189--285}, [\href{http://arxiv.org/abs/0901.0002}{{\tt
  0901.0002}}].

\bibitem{CyrRacine:2012fz}
F.-Y. Cyr-Racine and K.~Sigurdson, \emph{{Cosmology of atomic dark matter}},
  \href{http://dx.doi.org/10.1103/PhysRevD.87.103515}{\emph{Phys. Rev.} {\bf
  D87} (2013) 103515}, [\href{http://arxiv.org/abs/1209.5752}{{\tt
  1209.5752}}].

\bibitem{Petraki:2014uza}
K.~Petraki, L.~Pearce and A.~Kusenko, \emph{{Self-interacting asymmetric dark
  matter coupled to a light massive dark photon}},
  \href{http://dx.doi.org/10.1088/1475-7516/2014/07/039}{\emph{JCAP} {\bf 1407}
  (2014) 039}, [\href{http://arxiv.org/abs/1403.1077}{{\tt 1403.1077}}].

\bibitem{Boddy:2016bbu}
K.~K. Boddy, M.~Kaplinghat, A.~Kwa and A.~H.~G. Peter, \emph{{Hidden Sector
  Hydrogen as Dark Matter: Small-scale Structure Formation Predictions and the
  Importance of Hyperfine Interactions}},
  \href{http://dx.doi.org/10.1103/PhysRevD.94.123017}{\emph{Phys. Rev.} {\bf
  D94} (2016) 123017}, [\href{http://arxiv.org/abs/1609.03592}{{\tt
  1609.03592}}].

\bibitem{Foot:2014uba}
R.~Foot and S.~Vagnozzi, \emph{{Dissipative hidden sector dark matter}},
  \href{http://dx.doi.org/10.1103/PhysRevD.91.023512}{\emph{Phys. Rev.} {\bf
  D91} (2015) 023512}, [\href{http://arxiv.org/abs/1409.7174}{{\tt
  1409.7174}}].

\bibitem{Abazajian:2013oma}
K.~N. Abazajian et~al., \emph{{Neutrino Physics from the Cosmic Microwave
  Background and Large Scale Structure}},
  \href{http://dx.doi.org/10.1016/j.astropartphys.2014.05.014}{\emph{Astropart.
  Phys.} {\bf 63} (2015) 66--80}, [\href{http://arxiv.org/abs/1309.5383}{{\tt
  1309.5383}}].

\bibitem{Vogel:2013raa}
H.~Vogel and J.~Redondo, \emph{{Dark Radiation constraints on minicharged
  particles in models with a hidden photon}},
  \href{http://dx.doi.org/10.1088/1475-7516/2014/02/029}{\emph{JCAP} {\bf 1402}
  (2014) 029}, [\href{http://arxiv.org/abs/1311.2600}{{\tt 1311.2600}}].

\bibitem{Angloher:2015ewa}
{\scshape CRESST} collaboration, G.~Angloher et~al., \emph{{Results on light
  dark matter particles with a low-threshold CRESST-II detector}},
  \href{http://dx.doi.org/10.1140/epjc/s10052-016-3877-3}{\emph{Eur. Phys. J.}
  {\bf C76} (2016) 25}, [\href{http://arxiv.org/abs/1509.01515}{{\tt
  1509.01515}}].

\bibitem{Gutlein:2014gma}
A.~G{\"u}tlein et~al., \emph{{Impact of coherent neutrino nucleus scattering on
  direct dark matter searches based on CaWO$_4$ crystals}},
  \href{http://dx.doi.org/10.1016/j.astropartphys.2015.03.010}{\emph{Astropart.
  Phys.} {\bf 69} (2015) 44--49}, [\href{http://arxiv.org/abs/1408.2357}{{\tt
  1408.2357}}].

\bibitem{Cline:2012is}
J.~M. Cline, Z.~Liu and W.~Xue, \emph{{Millicharged Atomic Dark Matter}},
  \href{http://dx.doi.org/10.1103/PhysRevD.85.101302}{\emph{Phys. Rev.} {\bf
  D85} (2012) 101302}, [\href{http://arxiv.org/abs/1201.4858}{{\tt
  1201.4858}}].

\bibitem{Cirelli:2016rnw}
M.~Cirelli, P.~Panci, K.~Petraki, F.~Sala and M.~Taoso, \emph{{Dark Matter's
  secret liaisons: phenomenology of a dark U(1) sector with bound states}},
  \href{http://arxiv.org/abs/1612.07295}{{\tt 1612.07295}}.

\bibitem{Bringmann:2016din}
T.~Bringmann, F.~Kahlhoefer, K.~Schmidt-Hoberg and P.~Walia, \emph{{Strong
  constraints on self-interacting dark matter with light mediators}},
  \href{http://arxiv.org/abs/1612.00845}{{\tt 1612.00845}}.

\bibitem{Panci:2014gga}
P.~Panci, \emph{{New Directions in Direct Dark Matter Searches}},
  \href{http://dx.doi.org/10.1155/2014/681312}{\emph{Adv. High Energy Phys.}
  {\bf 2014} (2014) 681312}, [\href{http://arxiv.org/abs/1402.1507}{{\tt
  1402.1507}}].

\bibitem{Delaunay:2006ws}
C.~Delaunay, C.~Grojean and G.~Servant, \emph{{The Higgs in the sky: Production
  of gravitational waves during a first-order phase transition}},
  \href{http://dx.doi.org/10.1063/1.2735127}{\emph{AIP Conf. Proc.} {\bf 903}
  (2007) 24--31}.

\bibitem{Vaskonen:2016yiu}
V.~Vaskonen, \emph{{Electroweak baryogenesis and gravitational waves from a
  real scalar singlet}},  \href{http://arxiv.org/abs/1611.02073}{{\tt
  1611.02073}}.

\bibitem{Huang:2016odd}
F.~P. Huang, Y.~Wan, D.-G. Wang, Y.-F. Cai and X.~Zhang, \emph{{Hearing the
  echoes of electroweak baryogenesis with gravitational wave detectors}},
  \href{http://dx.doi.org/10.1103/PhysRevD.94.041702}{\emph{Phys. Rev.} {\bf
  D94} (2016) 041702}, [\href{http://arxiv.org/abs/1601.01640}{{\tt
  1601.01640}}].

\bibitem{Crowder:2005nr}
J.~Crowder and N.~J. Cornish, \emph{{Beyond LISA: Exploring future
  gravitational wave missions}},
  \href{http://dx.doi.org/10.1103/PhysRevD.72.083005}{\emph{Phys. Rev.} {\bf
  D72} (2005) 083005}, [\href{http://arxiv.org/abs/gr-qc/0506015}{{\tt
  gr-qc/0506015}}].

\bibitem{Kudoh:2005as}
H.~Kudoh, A.~Taruya, T.~Hiramatsu and Y.~Himemoto, \emph{{Detecting a
  gravitational-wave background with next-generation space interferometers}},
  \href{http://dx.doi.org/10.1103/PhysRevD.73.064006}{\emph{Phys. Rev.} {\bf
  D73} (2006) 064006}, [\href{http://arxiv.org/abs/gr-qc/0511145}{{\tt
  gr-qc/0511145}}].

\bibitem{Kawamura:2006up}
S.~Kawamura et~al., \emph{{The Japanese space gravitational wave antenna
  DECIGO}}, \href{http://dx.doi.org/10.1088/0264-9381/23/8/S17}{\emph{Class.
  Quant. Grav.} {\bf 23} (2006) S125--S132}.

\bibitem{Carrington:1991hz}
M.~E. Carrington, \emph{{The Effective potential at finite temperature in the
  Standard Model}},
  \href{http://dx.doi.org/10.1103/PhysRevD.45.2933}{\emph{Phys. Rev.} {\bf D45}
  (1992) 2933--2944}.

\bibitem{Carena:2004ha}
M.~Carena, A.~Megevand, M.~Quiros and C.~E.~M. Wagner, \emph{{Electroweak
  baryogenesis and new TeV fermions}},
  \href{http://dx.doi.org/10.1016/j.nuclphysb.2005.03.025}{\emph{Nucl. Phys.}
  {\bf B716} (2005) 319--351}, [\href{http://arxiv.org/abs/hep-ph/0410352}{{\tt
  hep-ph/0410352}}].

\bibitem{Curtin:2016urg}
D.~Curtin, P.~Meade and H.~Ramani, \emph{{Thermal Resummation and Phase
  Transitions}},  \href{http://arxiv.org/abs/1612.00466}{{\tt 1612.00466}}.

\end{thebibliography}

\providecommand{\href}[2]{#2}\begingroup\raggedright\endgroup

\end{document}